\documentclass[journal]{IEEEtran}
\usepackage{amsmath,amssymb,amsfonts,amsthm,mathrsfs,bm}
\usepackage{algorithm}
\usepackage{algpseudocodex}
\usepackage{array}
\usepackage[caption=false,font=normalsize,labelfont=sf,textfont=sf]{subfig}
\usepackage{textcomp}
\usepackage{stfloats}
\usepackage{url}
\usepackage{booktabs}
\usepackage{verbatim}
\usepackage{graphicx}
\usepackage{xcolor}
\usepackage[numbers,sort&compress]{natbib}
\usepackage{nicefrac}
\newcommand{\Fig}{Fig. }
\newcommand{\Alg}{Algorithm }
\newcommand{\Sec}{Section }
\newcommand{\Tab}{Table }
\newcommand{\Appdx}{Appendix }
\newcommand{\Thm}{Theorem }
\newcommand{\Assump}{Assumption }

\newtheorem{theorem}{Theorem}
\newtheorem{assumption}{Assumption}

\newcommand{\fourier}[1]{\mathscr{F}\{ #1 \}}
\newcommand{\cosTrans}[1]{\mathscr{F}_c\{ #1 \}}

\newcommand{\reals}{\mathbb{R}}
\newcommand{\complex}{\mathbb{C}}
\newcommand{\nnReals}{\reals_{\geq 0}}
\newcommand{\posReals}{\reals_{> 0}}

\newcommand{\sinc}{\text{sinc}}

\newcommand{\vectorize}{\text{vec}}
\newcommand{\unvectorize}[2]{\text{unvec}_{#1, #2}}

\newcommand{\f}{f}
\newcommand{\g}{g}
\newcommand{\coord}{\mathbf{x}}
\newcommand{\coordOneD}{x}
\newcommand{\freqCoord}{\boldsymbol\omega}
\newcommand{\freqCoordOneD}{\omega}
\newcommand{\vecInd}{k}
\newcommand{\coordElem}{\coordOneD_\vecInd}
\newcommand{\freqCoordElem}{\freqCoordOneD_\vecInd}
\newcommand{\fDims}{n}
\newcommand{\iUnit}{i}
\newcommand{\Fc}{F}
\newcommand{\Gc}{G}
\newcommand{\contHessMinDom}{\Omega} 
\newcommand{\signFuncSet}{\mathcal{S}}
\newcommand{\signFunc}{s}
\newcommand{\signFuncSol}{\Hat{\signFunc}}
\newcommand{\hess}{H}
\newcommand{\disF}{\mathsf{f}}
\newcommand{\disG}{\mathsf{g}}
\newcommand{\disSign}{\mathsf{s}}
\newcommand{\imM}{m}
\newcommand{\imN}{n}
\newcommand{\imIndI}{i}
\newcommand{\imIndJ}{j}
\newcommand{\imInds}{i,j}
\newcommand{\disHess}{\mathsf{H}}
\newcommand{\disHessXX}{\disHess^{xx}}
\newcommand{\disHessXY}{\disHess^{xy}}
\newcommand{\disHessYY}{\disHess^{yy}}
\newcommand{\signVec}{\mathbf{\signFunc}}
\newcommand{\G}{G}
\newcommand{\centFirstDiffMat}[1]{{D^1_{ #1 }}}
\newcommand{\centSecDiffMat}[1]{{D^2_{ #1 }}}
\newcommand{\Id}[1]{I_{#1}}
\newcommand{\hessDiffMatXX}{D^{xx}}
\newcommand{\hessDiffMatXY}{D^{xy}}
\newcommand{\hessDiffMatYY}{D^{yy}}
\newcommand{\norm}[1]{\left\Vert #1 \right\Vert}
\newcommand{\Qs}{\Tilde{Q}}
\newcommand{\qTilde}{q}
\newcommand{\qVecTilde}{\mathbf{q}}
\newcommand{\QTilde}{Q}

\newcommand{\bVec}{\mathbf{b}}
\newcommand{\bVecOpt}{\bVec^*}
\newcommand{\onesVec}{\mathbf{1}}
\newcommand{\cVec}{\mathbf{c}}
\newcommand{\weights}{\mathsf{w}}
\newcommand{\subblockM}{{m'}}
\newcommand{\subblockN}{{n'}}
\newcommand{\blockedM}{M}
\newcommand{\blockedN}{N}
\newcommand{\energyQuantile}{\alpha}
\newcommand{\algIter}{k}
\newcommand{\isEnergetic}{\mathsf{v}}
\newcommand{\candidateCells}{\mathsf{c}}
\newcommand{\neighborhood}{\mathcal{N}}
\newcommand{\nodeI}{l}
\newcommand{\nodeJ}{{l'}}
\newcommand{\blkIndI}{i}
\newcommand{\blkIndJ}{j}
\newcommand{\sHat}{\Hat{\mathsf{s}}}
\newcommand{\bordWidth}{\beta}
\newcommand{\ints}{\mathbb{Z}}
\newcommand{\nnInts}{\ints_{\geq0}}
\newcommand{\signSol}{\Hat{\disSign}}
\newcommand{\matElement}[3]{{#1}_{#2,#3}}
\newcommand{\matBlk}[3]{{#1}^{#2,#3}}
\newcommand{\signInterim}{\Tilde{\disSign}}
\newcommand{\gInterim}{\Tilde{\disG}}
\newcommand{\signBlkMat}{\overline{\disSign}}
\newcommand{\T}{T}

\newcommand{\signBlkVec}{\overline{\signVec}}
\newcommand{\preDctIm}{\mathsf{F}}
\newcommand{\image}{\preDctIm_0}
\newcommand{\imPadded}{\preDctIm_p}
\newcommand{\orgM}{m_0}
\newcommand{\orgN}{n_0}
\newcommand{\padM}{m_p}
\newcommand{\padN}{n_p}
\newcommand{\DctMat}[1]{C_{#1}}
\newcommand{\recF}{\Hat{\disF}}
\newcommand{\recPadIm}{\Hat{\preDctIm}}
\newcommand{\recIm}{\recPadIm_0}
\newcommand{\correctnessMat}{\mathsf{e}}
\newcommand{\neighborIndsSet}{\mathcal{N}}
\newcommand{\validNeighs}{\mathcal{V}}
\newcommand{\nss}{\text{NSS}}
\newcommand{\equalIndicator}{\mathbb{I}}
\newcommand{\imageSet}{\mathcal{I}}
\newcommand{\degenLocus}[1]{\Sigma( #1 )}
\newcommand{\degenImageSet}{\mathcal{D}_\contHessMinDom}
\newcommand{\perturb}{p}
\newcommand{\pertParam}{c}
\newcommand{\pertParams}{\mathbf{\pertParam}}
\newcommand{\pc}{\perturb_\pertParams}
\newcommand{\pertComp}{q}
\newcommand{\shift}{a}
\newcommand{\radius}{r}
\newcommand{\jet}{J}
\newcommand{\Pc}{P_\pertParams}
\newcommand{\jetc}{\jet_\pertParams}

\begin{document}

\title{Binary Phase Retrieval of Cosine Transforms \\ via Local Curvature Minimization}

\author{Ronald Ogden,~\IEEEmembership{Student Member,~IEEE,} Shwetadwip Chowdhury,~\IEEEmembership{Member,~IEEE,}\\ Takashi Tanaka,~\IEEEmembership{Senior Member,~IEEE,}
David Fridovich-Keil,~\IEEEmembership{Senior Member,~IEEE}
\thanks{Ronald Ogden (corresponding author) is with the Department of Aerospace Engineering and Engineering Mechanics at the University of Texas at Austin, Austin, TX (email: ronnieogden@utexas.edu).}
\thanks{Shwetadwip Chowdhury is with the Department of Electrical and Computer Engineering at the University of Texas at Austin, Austin, TX (email: shwetadwip.chowdhury@utexas.edu).}
\thanks{Takashi Tanaka is with the Department of Aeronautics and Astronautics and the Department of Electrical and Computer Engineering at Purdue University, West Lafayette, IN (email: tanaka16@purdue.edu).}
\thanks{David Fridovich-Keil is with the Department of Aerospace Engineering and Engineering Mechanics at the University of Texas at Austin, Austin, TX (email: dfk@utexas.edu).}}



\maketitle

\begin{abstract}
Cosine transforms see frequent use in image and video compression due to their ease of computation and high energy compaction. This has sparked interest within the optics community in computing cosine transforms optically for compression. When imaging in the Fourier plane to capture a cosine transform, one encounters a phase retrieval problem: optical fields have both magnitude and phase, but cameras only capture the magnitude of the field. Cosine transforms restrict the phase retrieval problem to a binary domain as opposed to the unit circle, but general phase retrieval solutions do not leverage this. In this work, we demonstrate that sign errors in phase retrieval for cosine transforms result in a significant increase in the magnitude of the Hessian of the transform at the error. Motivated by this, we develop an algorithm to solve this binary phase retrieval problem by minimizing the curvature of the reconstructed cosine transform. We demonstrate via computational experiments that given the magnitude of the cosine transform of an image, we can consistently reconstruct the original image within a multiscale structural similarity index of 0.95. We show that the solve time grows approximately linearly with the number of pixels solved. 
\end{abstract}


\section{Introduction}
\IEEEPARstart{F}{ourier} transforms have a plethora of applications in optics and signal processing, including compression. However, the closely related cosine transform sees significantly more use in image and video compression. For example, discrete cosine transforms (DCT) are used in the JPEG standard for image compression and the MPEG and H.261 standards for video compression \cite{sayood2017introduction}. Cosine transforms have become widely adopted due to their ease of computation (they can be computed as simply as the Fourier transform) as well as their energy compaction properties (having performance very similar to that of the optimal Karhunen-Loéve transform) \cite{ahmed1974discrete}.

The compressive benefits of cosine transforms have sparked a branch of research in Fourier optics centered around optically computed cosine transforms. \citet{gu1981optical} performed some of the earliest work on optically computed cosine transforms, experimentally demonstrating the benefits of cosine transforms from a rate distortion perspective. \citet{wong1992optical} proposed an optical method of processing video to meet the H.261 compression standard. \citet{ma2019sar} implemented an optical cosine transform system to perform synthetic aperture radar image compression. \citet{ogden2026rate} proposed utilizing optically computed cosine transforms with event cameras to perform high speed visual scene compression. 

Optical sensing devices typically capture the magnitude of an optical field, but not the phase, which is essential information in most applications, including those above. Except for \cite{gu1981optical}, each of these applications propose a beam-flooding technique for phase retrieval. There, a reference beam is superimposed on the optical cosine transform such that the resulting field is converted from a real-valued field to a nonnegative-valued field. If one has an image of the reference beam, an image of the standalone cosine transform and the image of the superposition, one can theoretically retrieve the phase of the cosine transform. However, the beam-flooding method relies on precise alignment to achieve interference sufficiently accurate to perform effective phase retrieval. This practically results in performance that is not robust to misalignment and other implementation errors, necessitating other more mature phase retrieval solutions.

General phase retrieval problems are posed as follows. Given measurements of the form
\begin{equation} \label{eq:genPhaseMeas}
    y_k = |\mathbf{a}_k^\top \mathbf{z}|^2,
\end{equation}
and known sampling vectors $\mathbf{a}_k\in\complex^n$ for $k\in[K]$, where $K$ is the number of measurements and $[K]:=\{1,2,3,...,K\}$, find the $\mathbf{z}\in\complex^n$ satisfying \eqref{eq:genPhaseMeas}. This problem has applications in crystallography, coherent imaging, astronomy, quantum mechanics, and acoustic processing, among many other fields. Many solutions exist for general phase retrieval problems, which can be readily applied to Fourier imaging. \citet{shechtman2015phase} provide a review of modern phase retrieval techniques in optics. Notable examples include alternating projection algorithms like the Gerchberg-Saxton algorithm \cite{gerchberg1994practical} or the hybrid input-output method \cite{fienup1982phase}, solving the transport of intensity equation \cite{zuo2020transport}, modern optimization techniques such as PhaseLift \cite{candes2013phaselift} or Wirtinger Flow \cite{candes2015phase} and a variety of learning-based approaches \cite{sinha2017lensless,goy2018low,zhang2024wrong,kang2020phase,metzler2018prdeep}. Each of these methods pose the phase retrieval problem as a decision over a complex domain. 

\begin{figure*}[!t]
\centering
\includegraphics[width=0.98\textwidth,trim=0 0 0 10.5cm, clip]{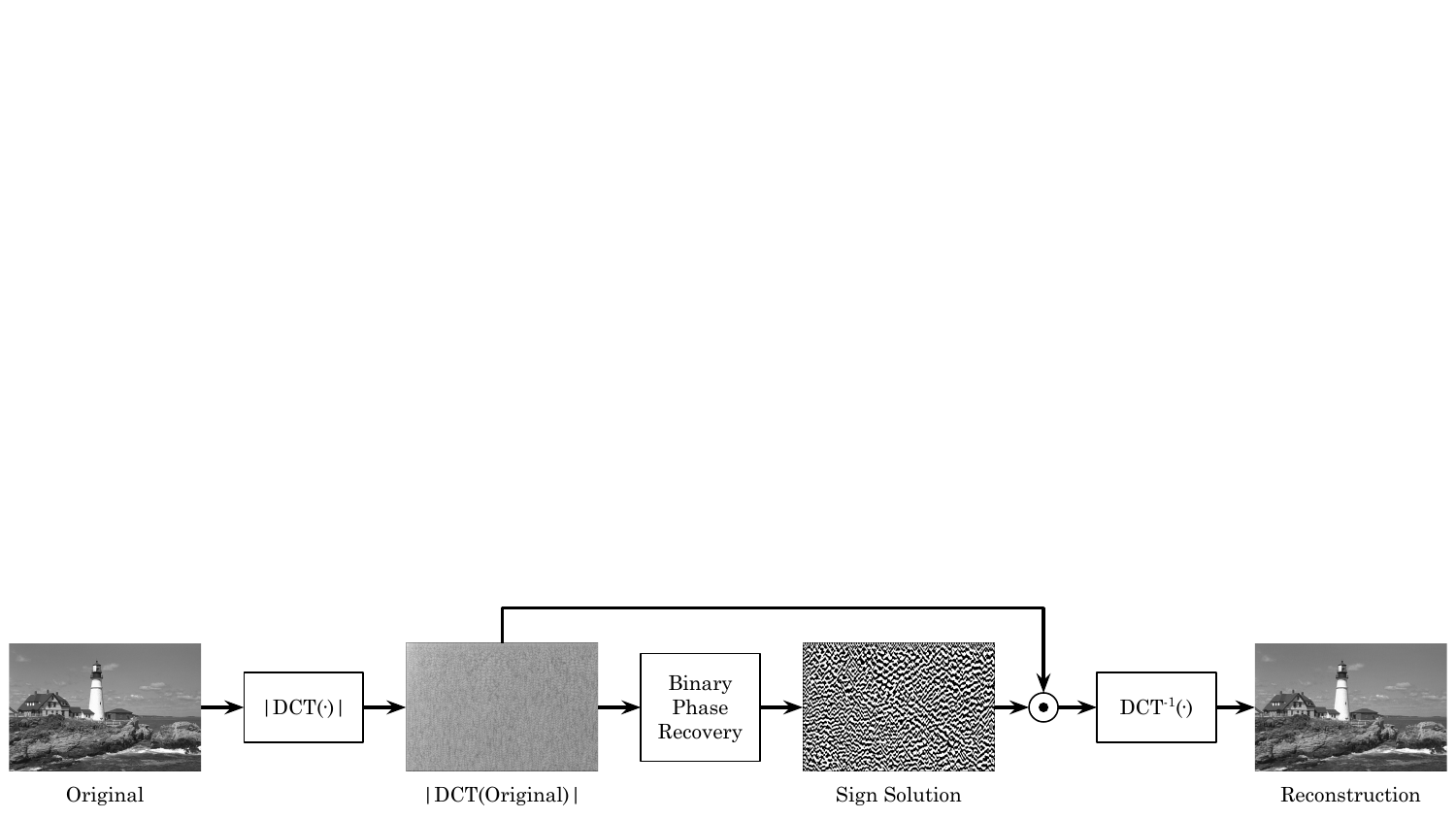}
\caption{Motivating application of the proposed phase retrieval method. Given access to the magnitude of the DCT of an image, we aim to recover the sign of the transform such that the original image can be reconstructed with high accuracy.}
\label{fig:compExp}
\end{figure*}

While the Fourier transform of a real-valued function can be complex-valued, the cosine transform of a real-valued function must be real-valued, allowing one to restrict the phase retrieval problem in the latter case to a binary domain as opposed to the full complex torus. State-of-the-art phase retrieval solutions cannot leverage this principle when applied to cosine transforms. In the context of image/video compression, the strong energy compaction of cosine transforms allows one to devote computational resources more efficiently during phase retrieval, focusing efforts in higher energy regions of the frequency domain. In this work, we leverage these advantages to develop a binary phase retrieval algorithm for cosine transforms of images. \Fig\ref{fig:compExp} depicts a block diagram of the binary phase recovery workflow.

\section{Phase Retrieval Methodology}

In this section we introduce fundamental properties of cosine transforms of images, motivate the Hessian norm minimization framework for binary phase retrieval of cosine transforms and convert the resulting problem into a standard-form binary quadratic program.
\subsection{Cosine Transforms of Images}\label{sec:cosTransProps}
Let the $\fDims$-dimensional Fourier transform of a function $\f:\reals^\fDims\mapsto\complex$ be defined as
\begin{equation}
    \fourier{\f}(\freqCoord) := \int_{\reals^\fDims}\f(\coord)e^{-2\pi\iUnit\coord\cdot\freqCoord} \,d\coord,
\end{equation}
and its $\fDims$-dimensional cosine transform be defined as
\begin{equation}
    \cosTrans{\f}(\freqCoord) := 2^\fDims\int_{\nnReals^\fDims}\f(\coord) \prod_{\vecInd = 1}^{\fDims} \cos(2\pi\coordElem\freqCoordElem)\, d\coord,
\end{equation}
where $\coordElem$ and $\freqCoordElem$ denote the $\vecInd^\text{th}$ elements of the vectors $\coord$ and $\freqCoord$, respectively. The cosine transform of $\f$ is the Fourier transform of the even part of $\f$.  

Our objective is to recover an image given the magnitude of its cosine transform. In order for this to be possible, we must have that the magnitude of an image's cosine transform is unique to that image. Herein, we shall consider an ``image" to be a function $\f:\reals^2\mapsto\reals$ that is nonnegative, integrable and has compact support. One can always shift an image so that its support lies in $\nnReals^2$ and then reflect the resulting function about each axis to produce an even function. The Fourier transform of the resulting function must be equivalent to its cosine transform. Given this definition, the intensity of the cosine transform of an image is unique. We state this property formally below.

\begin{theorem} \label{thm:uniqueness}
    Suppose $\f,\g:\reals^2\mapsto\reals$ are both even, nonnegative, integrable functions with compact support that satisfy $|\cosTrans{\f}(\freqCoord)| = |\cosTrans{\g}(\freqCoord)|$ for all $\freqCoord\in\reals^2$. Then $\f(\coord) = \g(\coord)$ almost everywhere.
\end{theorem}

\Appdx\ref{appdx:uniquenessProof} proves \Thm\ref{thm:uniqueness}. Similar results exist for many variations on the phase retrieval problem. For example, \citet{hayes1982reconstruction} analyzed the conditions that imply the uniqueness of a discrete multidimensional signal given the magnitude or phase of its Fourier transform. See Table 1 in \cite{shechtman2015phase} for the set of conditions necessary and/or sufficient to imply unique phase retrieval solutions in various versions of the phase retrieval problem.

In general phase retrieval, such uniqueness is not always guaranteed. In the case where one wishes to retrieve the phase of a Fourier transform given its magnitude, there are three ``trivial ambiguities" referred to in literature \cite{shechtman2015phase}, among other potential ambiguities:
\begin{enumerate}
    \item global phase shift: $\Tilde{\f}(\coord) = \f(\coord)e^{\iUnit\freqCoord_0}$,
    \item conjugate inversion: $\Tilde{\f}(\coord) = \f(-\coord)^*$,
    \item spatial shift: $\Tilde{\f}(\coord) = \f(\coord-\coord_0)$.
\end{enumerate}
The Fourier transform and cosine transform of any even function are equivalent, so \Thm\ref{thm:uniqueness} precludes these ambiguities in our framework. We examine each ambiguity to provide some intuition behind its resolution. We defined images to be real and nonnegative, immediately removing the first ambiguity. Extending a real image to be even removes the second ambiguity. The triviality of this third ambiguity may depend on the specific application. If one aims to optically encode the movement of an object by imaging the frequency domain as done in \cite{ogden2026rate}, one needs to ensure spatial shifts affect the magnitude of the Fourier transform. If $\Tilde{\f}(\coord) = \f(\coord-\coord_0)$, then 
\begin{equation}
    \fourier{\Tilde{\f}}(\freqCoord) = \fourier{\f}(\freqCoord)e^{-2\pi\iUnit\coord_0\cdot\freqCoord},
\end{equation}
which differs from $\fourier{\f}$ by a phase ramp. However, $|\fourier{\Tilde{\f}}|^2$ is equivalent to $|\fourier{\f}|^2$. In contrast, cosine transforms can capture space-domain shifts in their intensity maps. As an illustration, consider the 1-dimensional example where $\f(\coordOneD) = 1$ for $\coordOneD\in [0,1]$ and $\f(\coordOneD) = 0$ otherwise, and $\Tilde{\f}(\coordOneD) = \f(\coordOneD-1)$. Then
\begin{align}
    \cosTrans{\f}(\freqCoordOneD) &= 2\sinc(2\freqCoordOneD), \\
    \cosTrans{\Tilde{\f}}(\freqCoordOneD) &= 4\sinc(4\freqCoordOneD) - 2\sinc(2\freqCoordOneD).
\end{align}
The intensity of these cosine transforms differ significantly. Thus, the intensity of the cosine transform can retain information about the original function that the intensity of the Fourier transform could not without the addition of another tool, such as a stationary reference object in the image domain.

\subsection{Motivating Hessian Norm Minimization}\label{sec:hessNormMot}
For ease of notation, we will use $\f$ to denote the cosine transform of an image in this section onward. Given $\f:\reals^2\mapsto\reals$, we aim to recover $\f$ on the domain $\contHessMinDom := [0,a]\times[0,b]$ up to a global sign ambiguity, given only its magnitude. In particular, given only $\g(\coord) = |\f(\coord)|$ with the initial condition $\text{sgn}(f(\mathbf{0})) = 1$, the objective is to find a map $\signFunc:\contHessMinDom\mapsto\{-1,1\}$ such that $\signFunc(\coord)\g(\coord) = \f(\coord)$ for all $\coord\in\contHessMinDom$. For arbitrary $\f$, this problem is ill-posed. However, one can leverage the following two assumptions which generally hold for the cosine transforms of natural images: 
\begin{assumption} \label{assump:contGrad}
    $\nabla\f$ is continuous\footnote{As noted in the proof of \Thm\ref{thm:uniqueness}, cosine transforms of images must be analytic and thus smooth, so \Assump\ref{assump:contGrad} must hold.} on $\contHessMinDom$.
\end{assumption}
\begin{assumption} \label{assump:noDegenCurve}
    There exists\footnote{See \Appdx\ref{appdx:nonDegeneracy} for a theoretical note of this assumption.} an $\varepsilon > 0$ such that $\Vert \nabla\f (\coord) \Vert \geq \varepsilon$ for almost all $\coord$ satisfying $\f(\coord) = 0$. 
\end{assumption}
These assumptions ensure that the zero level sets of $g$ serve as unambiguous boundaries between the domain of $f$ that is positively valued and that which is negatively valued. If these two assumptions hold, then when reconstructing $f$, if an incorrect sign change is assigned across a closed curve, the resultant reconstruction will have a discontinuity in its gradient along the curve and a Dirac delta in its Hessian. When integrating the norm of the Hessian over the domain, these delta functions can significantly add to the total value, making them a strong indicator that an incorrect sign has been assigned to a region.

\begin{figure}[!t]
\centering
\includegraphics[width=0.48\textwidth]{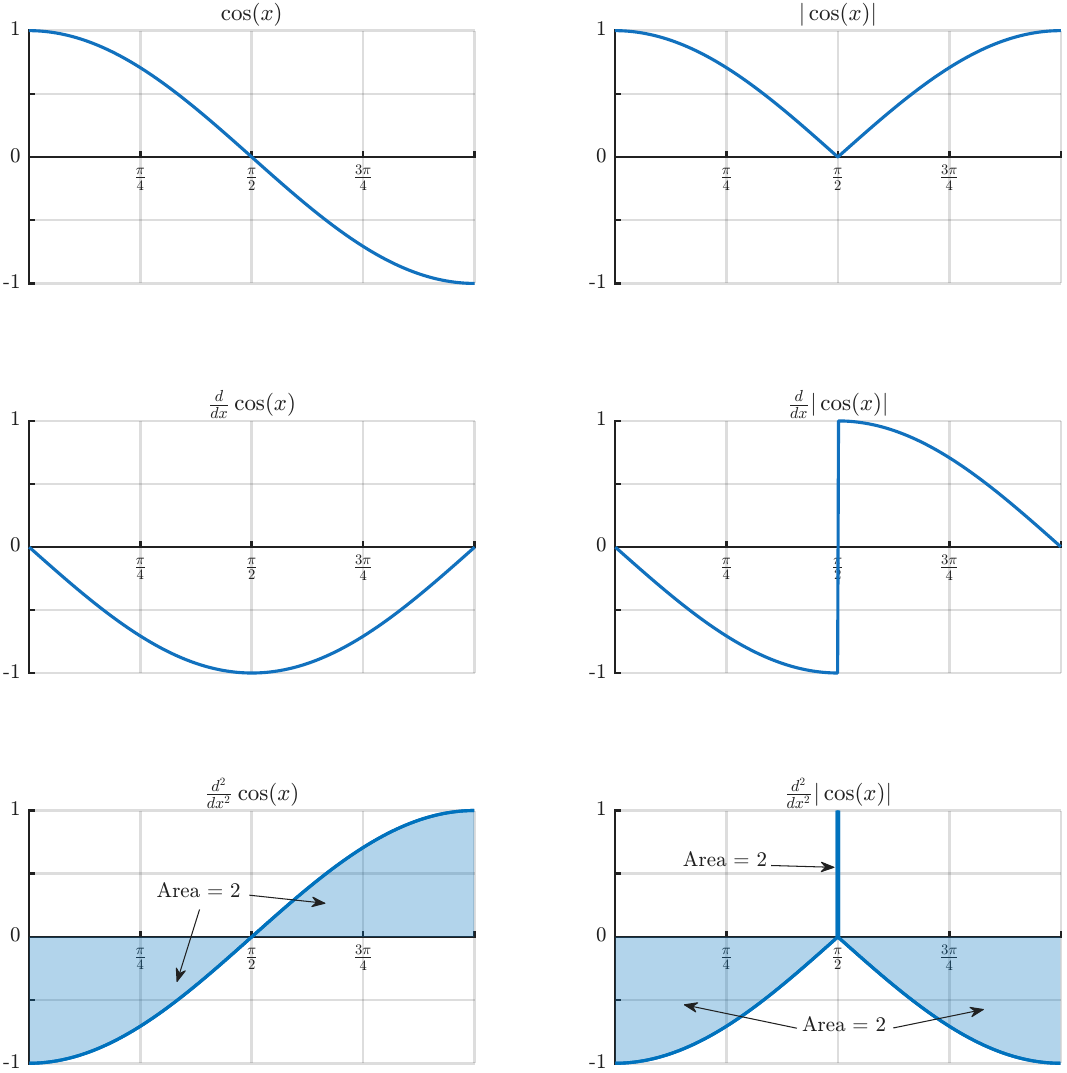}
\caption{Illustration of the impact of erroneous sign changes on the Hessian of a function. The left three plots illustrate a function satisfying \Assump\ref{assump:contGrad} and \Assump\ref{assump:noDegenCurve} with its first and second derivatives and the right three plots illustrate corresponding plots for an incorrect reconstruction.}
\label{fig:minHess}
\end{figure}

Consider the following illustrative 1-dimensional example of this phenomenon, depicted in \Fig\ref{fig:minHess}. Let $\f(\coordOneD) = \cos(\coordOneD)$ for $x\in[0,\pi]$, giving $\g(\coordOneD) = |\cos(\coordOneD)|$. The correct sign function is 
\begin{equation}
    \signFunc(\coordOneD) = \begin{cases}
        1,  &\coordOneD\in[0,\nicefrac{\pi}{2}), \\
        -1,  &\coordOneD\in(\nicefrac{\pi}{2},1].
    \end{cases}
\end{equation}
We shall compare this function to an incorrect sign function, $\signFuncSol(\coordOneD):=1$ for all $x\in[0,\pi]$. In the case of a correct sign function, the reconstruction of $\f$, that is $\signFunc\g$, has smooth first and second derivatives. Integrating the norm of the second derivative over $x\in[0,\pi]$ gives
\begin{align}
    \int_0^\pi \left\vert \frac{d^2}{d\coordOneD^2} \big(\signFunc(\coordOneD)\g(\coordOneD)\big)\right\vert \, d\coordOneD &= \int_0^\pi \left\vert\frac{d^2}{d\coordOneD^2} \cos(\coordOneD)\right\vert \, d\coordOneD \\
    &= \int_0^\pi \left\vert \cos(\coordOneD)\right\vert \, d\coordOneD \\
    &= 2.
\end{align}
In the case of an incorrect sign function, the reconstruction of $\f$, that is $\signFuncSol\g$, has a discontinuous first derivative and Dirac delta function in its second derivative. Integrating the norm of the second derivative over $x\in[0,\pi]$ gives
\begin{align}
    \int_0^\pi \left\vert \frac{d^2}{d\coordOneD^2} \big(\signFuncSol(\coordOneD)\g(\coordOneD)\big)\right\vert \, d\coordOneD &= \int_0^\pi \left\vert\frac{d^2}{d\coordOneD^2} |\cos(\coordOneD)|\right\vert \, d\coordOneD \\
    &= \int_0^\pi \left(\left\vert \cos(\coordOneD)\right\vert + 2\delta\left(\coordOneD-\nicefrac{\pi}{2}\right)\right) \, d\coordOneD \\
    &= 4.
\end{align}
The resulting integral is double that for the case of a correct sign function, demonstrating how the norm of the Hessian serves as an indicator for incorrect sign changes. 

Let $\signFuncSet$ denote the set of functions from $\contHessMinDom$ to $\{-1,1\}$, $\hess(f)$ denote the Hessian of a function $f$ and $\Vert \cdot \Vert_F$ denote the Frobenius norm of a matrix. The preceding discussion motivates us to solve the binary phase retrieval problem via the following optimization problem:
\begin{equation}\label{eq:contHessMin}
    \min_{\signFunc\in\signFuncSet} \int_{\contHessMinDom} \Vert \hess(\signFunc \g)(\coord) \Vert_F^2 \, d\coord.
\end{equation}

\subsection{Discretized Optimization Problem}\label{sec:discOptProb}
While our mathematical motivation has been presented in a continuous domain, in practice, we seek to recover the sign for functions measured discretely. Given sufficiently high spatial sampling of the function, the same motivation holds in the discrete analog. Let $\disF\in\reals^{\imM\times\imN}$ represent a discretization of $f$, and let $\disG\in\nnReals^{\imM\times\imN}$ be defined such that $\matElement{\disG}{\imIndI}{\imIndJ} = |\matElement{\disF}{\imIndI}{\imIndJ}|$ for all $\imIndI\in[\imM]$ and $\imIndJ\in[\imN]$. In order to formulate \eqref{eq:contHessMin} discretely, we utilize the following central difference approximation of the Hessian, $\disHess(\disF)$:
\begin{equation}
    \matElement{\disHess(\disF)}{\imIndI}{\imIndJ} := \begin{pmatrix}
        \matElement{\disHessXX(\disF)}{\imIndI}{\imIndJ} & \matElement{\disHessXY(\disF)}{\imIndI}{\imIndJ} \\
        \matElement{\disHessXY(\disF)}{\imIndI}{\imIndJ} & \matElement{\disHessYY(\disF)}{\imIndI}{\imIndJ}
    \end{pmatrix},
\end{equation}
where $\disHessXX(\disF)$, $\disHessXY(\disF)$, $\disHessYY(\disF) \in \reals^{\imM\times\imN}$ are matrices with entries given by
\begin{align}
    \matElement{\disHessXX(\disF)}{\imIndI}{\imIndJ} &:= \matElement{\disF}{\imIndI-1}{\imIndJ} -2\matElement{\disF}{\imIndI}{\imIndJ} + \matElement{\disF}{\imIndI+1}{\imIndJ}, \label{eq:hessDefXX} \\
    \begin{split}
    \matElement{\disHessXY(\disF)}{\imIndI}{\imIndJ} &:= (\matElement{\disF}{\imIndI-1}{\imIndJ-1} -\matElement{\disF}{\imIndI+1}{\imIndJ-1} \\
    &\qquad -\matElement{\disF}{\imIndI-1}{\imIndJ+1} + \matElement{\disF}{\imIndI+1}{\imIndJ+1})/4, 
    \end{split} \label{eq:hessDefXY} \\
    \matElement{\disHessYY(\disF)}{\imIndI}{\imIndJ} &:= \matElement{\disF}{\imIndI}{\imIndJ-1} -2\matElement{\disF}{\imIndI}{\imIndJ} + \matElement{\disF}{\imIndI}{\imIndJ+1} \label{eq:hessDefYY}
\end{align}
for all $\imIndI\in[\imM]$ and $\imIndJ\in[\imN]$. We adopt a same-padding scheme to approximate the Hessian at the border of $\disF$. That is, for $\disF\in\reals^{\imM\times\imN}$, we define $\disF_{0,\imIndJ}:=\disF_{1,\imIndJ}$, $\matElement{\disF}{\imM+1}{\imIndJ}:=\matElement{\disF}{\imM}{\imIndJ}$ for all $\imIndJ\in\{0,1,...,\imN+1\}$ and $\matElement{\disF}{\imIndI}{0}:=\matElement{\disF}{\imIndI}{1}$, $\matElement{\disF}{\imIndI}{\imN+1}:=\matElement{\disF}{\imIndI}{\imN}$ for all $\imIndI\in\{0,1,...,\imM+1\}$. With this approximation, the new optimization problem becomes
\begin{equation}\label{eq:disHessMin}
    \min_{\disSign\in\{-1,1\}^{m\times n}} \sum_{\imInds} \Vert \matElement{\disHess(\disSign \odot\disG)}{\imIndI}{\imIndJ} \Vert_F^2,
\end{equation}
where $\disSign\odot\disG$ denotes the Hadamard product of $\disSign$ and $\disG$.

\subsection{Binary Quadratic Program Formulation} \label{sec:quboForm}

All terms in the objective function in \eqref{eq:disHessMin} are quadratic expressions with respect to elements of $\disSign$, so \eqref{eq:disHessMin} is a binary quadratic program. In this section, we reformulate \eqref{eq:disHessMin} in a classical form recognizable by standard solvers such as Gurobi \cite{gurobi} or Mosek \cite{mosek}. 

Suppose $\disSign$ has columns $\signVec_1$, $\signVec_2$, ..., $\signVec_\imN$. Let $\signVec := \vectorize(\disSign) := (\signVec_1^\top, \signVec_2^\top, ..., \signVec_\imN^\top)^\top\in\{-1,1\}^{\imM\imN}$ and $\G := \text{diag}(\vectorize(\disG))\in\nnReals^{\imM\imN\times\imM\imN}$. Then  $\vectorize(\disSign\odot\disG) = \G\signVec$. Let $\centFirstDiffMat{\imM}$ represent the matrix that computes the central difference first derivative approximation on a vector of size $m$ and $\centSecDiffMat{\imM}$ represent the matrix that computes the central difference second derivative approximation on a vector of size $m$, both using same padding at the boundaries. Specifically,
\begin{align}
    \centFirstDiffMat{\imM} &:= \frac{1}{2}\begin{pmatrix}
        -1 & 1 &  &  & 0 \\
        -1 & 0 & 1 &  &  \\
         & -1 & \ddots & \ddots &  \\
         &  & \ddots & 0 & 1 \\
        0 &  &  & -1 & 1
    \end{pmatrix}, \\
    \centSecDiffMat{\imM} &:= \begin{pmatrix}
        -1 & 1 &  &  & 0 \\
         1 & -2 & 1 &  &  \\
          & 1 & \ddots & \ddots &  \\
          &  & \ddots & -2 & 1 \\
         0 &  &  & 1 & -1
    \end{pmatrix}.
\end{align}
Given these definitions we re-express \eqref{eq:hessDefXX} to \eqref{eq:hessDefYY} as
\begin{align}
    \disHessXX(\disF) &= \disF\centSecDiffMat{\imN}, \\
    \disHessXY(\disF) &= \centFirstDiffMat{\imM}\disF\centFirstDiffMat{\imN}^\top, \\
    \disHessYY(\disF) &= \centSecDiffMat{\imM}\disF.
\end{align}
As provided in \cite{petersen2008matrix}, given arbitrary matrices $A$, $X$, $B$ of compatible dimensions, the following relation holds:
\begin{equation} \label{eq:vectorizeProduct}
    \vectorize(AXB) = \left(B^\top \otimes A \right)\vectorize(X).
\end{equation}
Thus, if we define $\Id{\imM}$ to be the $\imM$-dimensional identity matrix and apply \eqref{eq:vectorizeProduct} to the preceding equations, we have that for arbitrary $\disF\in\reals^{\imM\times\imN}$,
\begin{align}
    \vectorize(\disHessXX(\disF)) &= \hessDiffMatXX \vectorize(\disF), \text{ where } \hessDiffMatXX:=\centSecDiffMat{\imN}\otimes\Id{\imM}, \label{eq:hessXX} \\
    \vectorize(\disHessXY(\disF)) &= \hessDiffMatXY \vectorize(\disF), \text{ where } \hessDiffMatXY:=\centFirstDiffMat{\imN}\otimes\centFirstDiffMat{\imM}, \label{eq:hessXY} \\
    \vectorize(\disHessYY(\disF)) &= \hessDiffMatYY \vectorize(\disF), \text{ where } \hessDiffMatYY:=\Id{\imN}\otimes\centSecDiffMat{\imM}, \label{eq:hessYY}
\end{align}
where $\otimes$ denotes the Kronecker product of two matrices.

Using the relationships introduced in this section, we have:
\begin{align}
    \lefteqn{\sum_{\imInds} \Vert \matElement{\disHess(\disSign \odot\disG)}{\imIndI}{\imIndJ} \Vert_F^2} \nonumber \\
    &= \sum_{\imInds} \Big( \vert \matElement{\disHessXX(\disSign \odot\disG)}{\imIndI}{\imIndJ} \vert^2 +2\vert \matElement{\disHessXY(\disSign \odot\disG)}{\imIndI}{\imIndJ} \vert^2 \nonumber \\
    &\qquad\qquad + \vert \matElement{\disHessYY(\disSign \odot\disG)}{\imIndI}{\imIndJ} \vert^2 \Big) \\
    &= \norm{\vectorize(\disHessXX(\disSign \odot\disG))}^2 + 2\norm{\vectorize(\disHessXY(\disSign \odot\disG))}^2 \nonumber \\
    &\qquad \qquad + \norm{\vectorize(\disHessYY(\disSign \odot\disG))}^2 \\
    &= \norm{\hessDiffMatXX\G\signVec}^2 + 2\norm{\hessDiffMatXY\G\signVec}^2 + \norm{\hessDiffMatYY\G\signVec}^2 \\
    &= \signVec^\top \G \left( (\hessDiffMatXX)^2 + 2{\hessDiffMatXY}^\top\hessDiffMatXY + (\hessDiffMatYY)^2 \right) \G \signVec.
\end{align}
Thus, the optimization problem can be expressed as 
\begin{equation}\label{eq:simpleOpt}
    \min_{\signVec\in\{-1,1\}^{\imM\imN}} \signVec^\top \Qs \signVec,
\end{equation}
where 
\begin{equation}\label{eq:simpleQDef}
\Qs := \G ( (\hessDiffMatXX)^2 + 2{\hessDiffMatXY}^\top\hessDiffMatXY + (\hessDiffMatYY)^2 ) \G.
\end{equation}
Note that \eqref{eq:simpleOpt} has at least two solutions due to the invariance of the objective to global sign changes. One can assume without loss of generality that the first element of $\signVec$ is 1. If $\Qs$ decomposes into block matrices as
\begin{equation}\label{eq:QDecomp}
    \Qs = \begin{pmatrix}
        \qTilde & \qVecTilde^\top \\
        \qVecTilde & \QTilde
    \end{pmatrix},
\end{equation}
where $\qTilde$ is a scalar, $\qVecTilde \in \reals^{\imM\imN-1}$ and $\QTilde\in\reals^{(\imM\imN-1)\times(\imM\imN-1)}$, then the last $\imM\imN-1$ elements of the minimizers of \eqref{eq:simpleOpt} are equivalent to the minimizer of following problem up to a global sign:
\begin{equation}\label{eq:noSignAmbOpt}
    \min_{\signVec\in\{-1,1\}^{\imM\imN-1}} \signVec^\top \QTilde \signVec + 2\qVecTilde^\top \signVec.
\end{equation}
Finally, we use the substitution $\signVec = 2\bVec-\onesVec$, where $\onesVec$ is a vector of all ones, to convert the decision variable from signs to binary digits. Thus, the final optimization problem takes the form
\begin{equation}\label{eq:binOpt}
    \min_{\bVec\in\{0,1\}^{\imM\imN-1}} \bVec^\top \QTilde\bVec + \cVec^\top \bVec,
\end{equation}
where $\cVec = \qVecTilde - \QTilde\onesVec$. This is a quadratic unconstrained binary optimization (QUBO) problem, which can be solved using off-the-shelf optimization software such as Gurobi \cite{gurobi} or Mosek \cite{mosek}. After computing a solution to \eqref{eq:binOpt}, one can recover a discrete sign function $\signSol\in\{-1,1\}^{\imM\times\imN}$ via
\begin{equation}
    \signSol = \unvectorize{\imM}{\imN}\begin{pmatrix}
        1 \\
        2\bVecOpt-\onesVec
    \end{pmatrix},
\end{equation}
where $\bVec^*$ is the solution to \eqref{eq:binOpt} and the $\unvectorize{\imM}{\imN}(\cdot)$ operator is the inverse of the $\vectorize(\cdot)$ operator introduced previously.

\section{Optimization Algorithm}\label{sec:optAlg}
QUBOs are known to be NP-hard problems in general \cite{punnen2022quadratic}. Solve times for such problems using exact techniques like branch-and-bound can grow exponentially with input size, posing an issue in this case where inputs consist of large arrays. In this section, we introduce \Alg\ref{alg:phaseRecovery}, which exploits the structure of cosine transforms of natural images to achieve effective binary phase retrieval quickly. \Alg\ref{alg:phaseRecovery} functions by
\begin{enumerate}
\item Dividing the array $\disG$ into blocks,
\item Determining a set of connected, high-energy blocks (measured as the squared Frobenius norm of the block), 
\item Solving \eqref{eq:binOpt} for each block in this high-energy set individually in parallel, and 
\item Resolving inter-block sign discrepancies via a related but far lower-dimensional QUBO. 
\end{enumerate}
The decomposition of $\disG$ enables the binary solver to operate on significantly smaller problems in parallel as opposed to attempting to search the intractably large domain of the original problem. By only solving the phase of high-energy blocks, \Alg\ref{alg:phaseRecovery} devotes its attention to recovering the phase of pixels that will most impact the reconstruction of the original image. The remainder of this section elaborates on the details of \Alg\ref{alg:phaseRecovery}. 

\begin{algorithm}
\caption{Binary Phase Retrieval}\label{alg:phaseRecovery}
\begin{algorithmic}[1] 

\Require 
Array $\disG\in\nnReals^{\imM\times\imN}$, block size $\subblockM\times\subblockN$, block inclusion ratio $\energyQuantile\in [0,1]$, block border width $\bordWidth\in\nnInts$.
\Ensure 
Sign (phase) solution $\sHat\in\{-1,0,1\}^{\imM\times\imN}$.

\State $(\imM,\imN) \gets \text{size}(\disG)$
\State $\blockedM\gets\imM/\subblockM$, $\blockedN\gets\imN/\subblockN$
\State $\matElement{\weights}{\blkIndI}{\blkIndJ} \gets \Vert \matBlk{\disG}{\blkIndI}{\blkIndJ} \Vert_F^2$ \textbf{for} $(\blkIndI,\blkIndJ)\in[\blockedM]\times[\blockedN]$

\State $\isEnergetic\gets$\Call{getEnergeticBlocks}{$\weights,\energyQuantile$} \Comment{\Alg\ref{alg:greedyMwcs}}
\State $\isEnergetic\gets$\Call{dilation}{$\isEnergetic$}

\For{$(\blkIndI,\blkIndJ)\in[\blockedM]\times[\blockedN]$} \Comment{Parallelizable loop}
    \If{$\matElement{\isEnergetic}{\blkIndI}{\blkIndJ} = 1$}
        \State $\matBlk{\signInterim}{\blkIndI}{\blkIndJ}_\bordWidth\gets\Call{solveBinaryPhase}{\matBlk{\disG}{\blkIndI}{\blkIndJ}_\bordWidth,1,1}$ \label{alg:line:parPhaseSolve}
        \State $\matBlk{\signInterim}{\blkIndI}{\blkIndJ}\gets\Call{trimBorder}{\matBlk{\signInterim}{\blkIndI}{\blkIndJ}_\bordWidth,\bordWidth}$
    \Else 
        \State $\matBlk{\signInterim}{\blkIndI}{\blkIndJ}\gets 0_{\subblockM\times\subblockN}$
    \EndIf
\EndFor
\State $\gInterim\gets\signInterim\odot\disG$
\State $\signBlkMat\gets\Call{solveBinaryPhase}{\gInterim,\subblockM,\subblockN}$ \label{alg:line:blockResolve}
\State $\signSol \gets \signInterim\odot\left(\signBlkMat\otimes\onesVec_{\subblockM\times\subblockN}\right)$ 
\State \Return $\signSol$

\Statex

\Function{solveBinaryPhase}{$\disG,\subblockM,\subblockN$}
\State $(\imM,\imN) \gets \text{size}(\disG)$
\State $\blockedM\gets\imM/\subblockM$, $\blockedN\gets\imN/\subblockN$
\State Compute $T$ via \eqref{eq:TDef} given $\subblockM,\subblockN,\blockedM,\blockedN$
\State Compute $\Qs$ via \eqref{eq:newQDef}
\State Extract $\QTilde$ and $\qVecTilde$ from $\Qs$ according to \eqref{eq:QDecomp}
\State $\cVec \gets \qVecTilde - \QTilde\onesVec$
\State $\bVecOpt \gets \arg\min_{\bVec\in\{0,1\}^{\blockedM\blockedN-1}} \bVec^\top \QTilde\bVec + \cVec^\top \bVec$
\State \Return $\unvectorize{\blockedM}{\blockedN}(1, \quad (2\bVecOpt-\onesVec)^\top)^\top$
\EndFunction

\end{algorithmic}
\end{algorithm}

\Alg\ref{alg:phaseRecovery} decomposes $\disG$ into blocks of size $\subblockM\times\subblockN$ with $\blockedM := \nicefrac{\imM}{\subblockM} \in \mathbb{N}$ and $\blockedN := \nicefrac{\imN}{\subblockN} \in \mathbb{N}$ encoding the number of blocks in each dimension. We denote by $\matBlk{\disG}{\blkIndI}{\blkIndJ}\in\nnReals^{\subblockM\times\subblockN}$ the $(\blkIndI,\blkIndJ)^\text{th}$ block of $\disG$, consisting of rows $(\blkIndI-1)\subblockM+1$ to $\blkIndI\subblockM$ and columns $(\blkIndJ-1)\subblockN+1$ to $\blkIndJ\subblockN$ of $\disG$.

One could solve for the phase of every block in $\disG$, but many blocks will have minimal impact when computing the IDCT of $\disG$. The energy compaction property of cosine transforms ensures that the majority of the energy in $\disG$ is concentrated in regions corresponding to low frequencies. \Alg\ref{alg:phaseRecovery} takes advantage of this, reducing the computational load by solving the phase of only the most energetic fraction of blocks, minimizing distortion upon image reconstruction. Given $\energyQuantile\in[0,1]$, a parameter denoting the fraction of blocks whose phase will be solved, \Alg\ref{alg:greedyMwcs} computes a mask, $\isEnergetic\in\{0,1\}^{\blockedM\times\blockedN}$, such that $\sum_{\imIndI,\imIndJ} \isEnergetic_{\imIndI,\imIndJ} = \lfloor \alpha \blockedM \blockedN \rfloor$. If $\matElement{\isEnergetic}{\blkIndI}{\blkIndJ} = 1$, then \Alg\ref{alg:phaseRecovery} will solve for the phase of block $\matBlk{\disG}{\blkIndI}{\blkIndJ}$. Otherwise, it will assign a sign of 0 to the corresponding block.\footnote{If an element of $\disSign$ is 0 instead of $\pm 1$, then $\disSign$ is infeasible with respect to the optimization problem in \eqref{eq:disHessMin}. This relaxation from the original formulation enables the removal of low energy frequencies during image reconstruction as opposed to arbitrary sign allocation.} Even under \Assump\ref{assump:contGrad} and \Assump\ref{assump:noDegenCurve}, binary phase retrieval up to a global sign is not possible on a function with a domain that is not connected, so \Alg\ref{alg:greedyMwcs} must ensure that the $1$-level set of $\isEnergetic$ is connected. 

\begin{figure}[!t]
\centering
\includegraphics[width=0.48\textwidth]{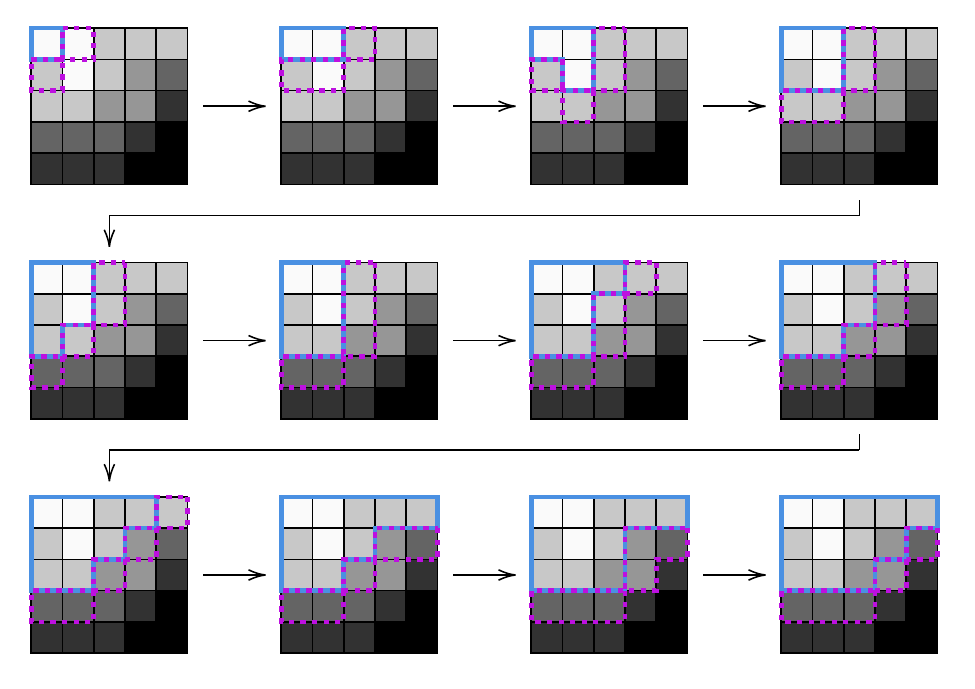}
\caption{Illustration of \Alg\ref{alg:greedyMwcs}. The energy of a block is depicted by its grayscale color (lighter corresponding to greater weights). The blocks included in the solution ($\isEnergetic$ in \Alg\ref{alg:greedyMwcs}) at each iteration are bounded by the solid blue border. The blocks considered as candidates ($\candidateCells$ in \Alg\ref{alg:greedyMwcs}) for the next iteration are bounded by a dotted magenta border.}
\label{fig:mwcs}
\end{figure}

Let $\weights\in\nnReals^{\blockedM\times\blockedN}$ given by $\matElement{\weights}{\blkIndI}{\blkIndJ} := \Vert\matBlk{\disG}{\blkIndI}{\blkIndJ}\Vert_F^2$ denote the array of block energies of $\disG$. From the discussion above, the mask $\isEnergetic$ must be connected and satisfy $\sum_{\imIndI,\imIndJ} \isEnergetic_{\imIndI,\imIndJ} = \lfloor \alpha \blockedM \blockedN \rfloor$ while aiming to maximize $\sum_{\imIndI,\imIndJ} \isEnergetic_{\imIndI,\imIndJ}\weights_{\imIndI,\imIndJ}$. \Alg\ref{alg:greedyMwcs} implements a greedy procedure to achieve this objective. \Alg\ref{alg:greedyMwcs} initializes the solution $\isEnergetic$ to contain only the highest energy block. At each iteration, the block with the highest energy that is adjacent to the current solution set is added. The array $\candidateCells\in\{0,1\}^{\blockedM\times\blockedN}$ in \Alg\ref{alg:greedyMwcs} indicates the blocks that are adjacent to the solution $\isEnergetic$ at a given iteration. \Fig\ref{fig:mwcs} illustrates \Alg\ref{alg:greedyMwcs} visually. \Alg\ref{alg:greedyMwcs} can result in a $\isEnergetic$ with only sparse connectivity, which can negatively impact binary phase retrieval. \Alg\ref{alg:phaseRecovery} applies a morphological dilation to $\isEnergetic$ to remedy this phenomenon.

\begin{algorithm}
\caption{Energy-Greedy Connected Block Masking}\label{alg:greedyMwcs}
\begin{algorithmic}[1] 

\Require 
Array of block energies $\weights\in\nnReals^{\blockedM\times\blockedN}$, block inclusion ratio $\energyQuantile\in [0,1]$.
\Ensure 
Indicator array $\isEnergetic\in\{0,1\}^{\blockedM\times\blockedN}$ of blocks in solution.
\State Initialize $\isEnergetic\gets 0_{\blockedM\times\blockedN}$,  $\candidateCells\gets 0_{\blockedM\times\blockedN}$
\State $\nodeI \gets \arg\max\weights$
\State $\isEnergetic_\nodeI \gets 1$
\State $\candidateCells_\nodeJ \gets 1$ \textbf{for} $\nodeJ\in\neighborhood(\nodeI)$ \Comment{$\neighborhood(\nodeI):=\{\nodeJ \text{ adjacent to } \nodeI\}$}

\For{$\algIter \gets 2:\lfloor \alpha \blockedM \blockedN \rfloor$}
    \State $\nodeI \gets \arg\max(\weights\odot\candidateCells)$
    \State $\isEnergetic_\nodeI \gets 1$, $\candidateCells_\nodeI \gets 0$
    
    \State \algorithmicfor\ $\nodeJ\in\neighborhood(\nodeI)$ \algorithmicif\ $\isEnergetic_\nodeJ \neq 1$ \algorithmicthen\ $\candidateCells_\nodeJ \gets 1$
    
\EndFor

\State \Return $\isEnergetic$

\end{algorithmic}
\end{algorithm}

In order to solve for the sign of an individual block, $\matBlk{\disG}{\blkIndI}{\blkIndJ}$, the block along with its bordering pixels are fed to the QUBO formulated in \Sec\ref{sec:quboForm} to solve for their sign. Let $\matBlk{\disG}{\blkIndI}{\blkIndJ}_\bordWidth$ denote the block augmented with a border of width $\bordWidth$, that is, the array consisting of rows $(\blkIndI-1)\subblockM+1-\bordWidth$ to $\blkIndI\subblockM+\bordWidth$ and columns $(\blkIndJ-1)\subblockN+1-\bordWidth$ to $\blkIndJ\subblockN+\bordWidth$ of $\disG$. Applying the procedure in \Sec\ref{sec:quboForm} to $\matBlk{\disG}{\blkIndI}{\blkIndJ}_\bordWidth$ will yield an output, $\matBlk{\signInterim}{\blkIndI}{\blkIndJ}_\bordWidth\in\{-1,1\}^{(\subblockM+\bordWidth)\times(\subblockN+\bordWidth)}$. Trimming the first $\bordWidth$ and last $\bordWidth$ rows and columns of $\matBlk{\signInterim}{\blkIndI}{\blkIndJ}_\bordWidth$ yields $\matBlk{\signInterim}{\blkIndI}{\blkIndJ}\in\{-1,1\}^{\subblockM\times\subblockN}$, the interim sign solution for the individual block. While we ultimately discard the solution for the sign of the border pixels, passing border pixels to the QUBO solver enables it to more accurately compute the Hessian and solve for the correct sign near the edge of a block. The width of the border $\bordWidth$ can be tuned to achieve the desired performance from the solver. Increasing $\bordWidth$ improves the accuracy of the solver at the cost of a solve time that grows exponentially with $\bordWidth$. Same padding is used whenever the border extends beyond the domain of the image.

\begin{figure}[!t]
\centering
\includegraphics[width=0.48\textwidth, trim=0 0 6cm 0, clip]{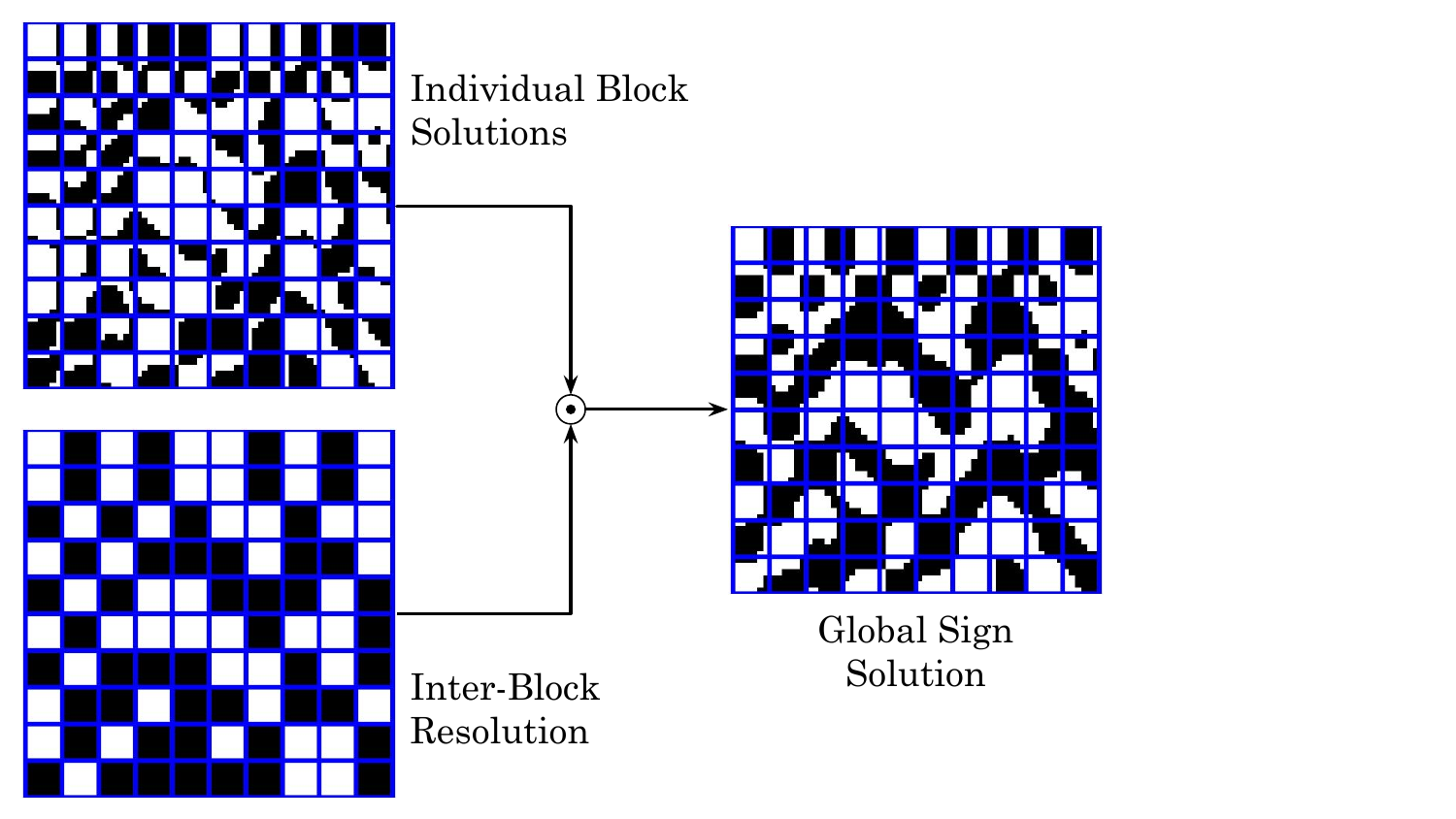}
\caption{Illustration of inter-block discrepancies and their resolution. The top-left image represents the individual block solutions concatenated as the interim solution, $\signInterim$. That image still contains inter-block sign discrepancies. The lower left image depicts the corresponding sign corrections necessary to resolve the discrepancies, $\signBlkMat\otimes\onesVec_{\subblockM\times\subblockN}$. The Hadamard product of these two arrays yields the full solution, $\signSol$.}
\label{fig:ibrDemo}
\end{figure}

The formulation in \eqref{eq:noSignAmbOpt} assumed that the first (top-left) pixel of the array undergoing phase retrieval has positive sign. While this assumption removes the sign solution ambiguity for a block, when applied to all of the individual blocks of an image, it leads to an inter-block sign discrepancy. The correct sign for the top-left pixel of every block will likely vary between blocks. See \Fig\ref{fig:ibrDemo} for an example of this phenomenon. Let $\signInterim\in\{-1,0,1\}^{\imM\times\imN}$ be the interim sign solution resulting from concatenating the individual block solutions, $\matBlk{\signInterim}{\blkIndI}{\blkIndJ}$, together (we include 0 as a potential element of $\signInterim$ to account for low-energy blocks assigned a value of 0 in the previous stage of \Alg\ref{alg:phaseRecovery}). In order to resolve this inter-block sign discrepancy, one can treat $\gInterim := \signInterim\odot\disG$ as a new function to recover the sign of.


We formulate this new phase retrieval problem as discussed in \Sec\ref{sec:quboForm}; however, here we define sign changes across entire blocks (rather than individual pixels) as the decision variables. In order to apply an array of signs $\signBlkMat\in\{-1,1\}^{\blockedM\times\blockedN}$ blockwise to $\gInterim$, that is, multiply the block $\matBlk{\gInterim}{\blkIndI}{\blkIndJ}$ by the sign $\matElement{\signBlkMat}{\blkIndI}{\blkIndJ}$ for all $\blkIndI\in[\blockedM]$ and $\blkIndJ\in[\blockedN]$, one can multiply $\gInterim$ elementwise by $\signBlkMat\otimes\onesVec_{\subblockM\times\subblockN}$, where $\onesVec_{\subblockM\times\subblockN}\in\reals^{\subblockM\times\subblockN}$ denotes a matrix of all ones. As shown in \Appdx\ref{sec:deriveB}, the following relationship holds:
\begin{equation}\label{eq:Trelation}    \vectorize(\signBlkMat\otimes\onesVec_{\subblockM\times\subblockN}) = \T\vectorize(\signBlkMat),
\end{equation}
where $T\in \{0,1\}^{\imM\imN\times\blockedM\blockedN}$ is given by
\begin{equation} \label{eq:TDef}
    \T := \Id{\blockedN}\otimes\onesVec_{\subblockN}\otimes\Id{\blockedM}\otimes\onesVec_{\subblockM} 
\end{equation}
and $\onesVec_{\subblockM}\in\reals^{\subblockM}$ denotes a vector of all ones. That is, $\T$ transforms a vector of blockwise signs to the corresponding vector of pixelwise signs. Using this relationship, to solve the Hessian minimization problem for $\gInterim$ as outlined in \Sec\ref{sec:quboForm}, we replace the definition for $\Qs$ in \eqref{eq:simpleQDef} with 
\begin{equation}\label{eq:newQDef}
\Qs:= \T^\top\G ( (\hessDiffMatXX)^2 + 2{\hessDiffMatXY}^\top\hessDiffMatXY + (\hessDiffMatYY)^2 ) \G\T
\end{equation}
and all instances of $\imM$ and $\imN$ with $\blockedM$ and $\blockedN$ respectively. In the case where $\subblockM=\subblockN=1$, we have $\blockedM = \imM$, $\blockedN = \imN$ and $T = \Id{\imM\imN}$, resulting in the definition of $\Qs$ in \eqref{eq:simpleQDef} and \eqref{eq:newQDef} being equivalent. If the resulting solution is denoted $\signBlkMat\in\{-1,1\}^{\blockedM\times\blockedN}$, then the final binary phase retrieval solution is given by
\begin{equation}
    \signSol = \signInterim\odot\left(\signBlkMat\otimes\onesVec_{\subblockM\times\subblockN}\right).
\end{equation}

One should note that this divide-and-conquer method of solving the QUBO over the whole image does not guarantee a globally optimal solution. However, as will be demonstrated in \Sec\ref{sec:compResults}, there remains little practical room for improvement in the quality of the phase retrieval. 

\section{Computational Results}\label{sec:compResults}
The performance of \Alg\ref{alg:phaseRecovery} was measured by testing it on the DCT magnitude of images in the Kodak Lossless True Color Image Suite \cite{franzen1999kodak}. This dataset consists of 24 images of size $768\times 512$ or $512\times 768$. \Fig\ref{fig:compExp} depicts the experimental pipeline for a single image. We compute the magnitude of the DCT of an image before feeding it to \Alg\ref{alg:phaseRecovery}. The resulting sign solution is applied to the DCT magnitude function to compute the reconstruction of the DCT. We take the inverse DCT of the result to compute the reconstructed image. As noted briefly in \Sec\ref{sec:discOptProb}, our motivation for Hessian norm minimization as a means to recover binary phase applies to the discrete case only if the DCT used has a sufficiently small sampling interval. We can create an arbitrarily small DCT sampling interval by padding the original image with zeros before taking the DCT. 

Given an input image $\image\in\nnReals^{\orgM\times\orgN}$ and nonnegative integer parameters $\padM$ and $\padN$, we pad $\image$ as follows:
\begin{equation}
    \imPadded = \begin{pmatrix}
        \image & 0_{\orgM\times\padN} \\
        0_{\padM\times\orgN} & 0_{\padM\times\padN}
    \end{pmatrix}.
\end{equation}
Then given the parameters defining the solver's block size $\subblockM$ and $\subblockN$, we compute $\imM = \subblockM\lfloor (\orgM+\padM)/\subblockM\rfloor$ and $\imN = \subblockN\lfloor (\orgN+\padN)/\subblockN\rfloor$ and construct $\preDctIm\in\reals^{\imM\times\imN}$ using the first $\imM$ rows and $\imN$ columns of $\imPadded$. This ensures that the dimensions of $\preDctIm$ are such that they divide evenly into blocks during phase retrieval. Let $\DctMat{\imM}\in\reals^{\imM\times\imM}$ be the matrix that performs the DCT-II transform on a vector:
\begin{equation}
    \matElement{(\DctMat{\imM})}{\imIndI}{\imIndJ} := \begin{cases}
        \sqrt{\frac{1}{\imM}}, &\imIndI = 1, \\
        \sqrt{\frac{2}{\imM}}\cos\frac{\pi(2\imIndI-1)(\imIndJ-1)}{2\imM}, &\imIndI\in [\imM]\setminus \{1\},
    \end{cases}
\end{equation}
for all $\imIndJ\in[\imN]$. Then we compute the two-dimensional DCT of $\preDctIm$:
\begin{equation}
    \disF = \DctMat{\imM}\preDctIm\DctMat{\imN}^\top.
\end{equation}
Letting $\disG$ be the element-wise magnitude of $\disF$, we apply \Alg\ref{alg:phaseRecovery} to $\disG$ with parameters $\subblockM$, $\subblockN$, $\energyQuantile$ and $\bordWidth$ to compute $\signSol$. We compute the reconstructed DCT as $\recF = \signSol\odot\disG$. Finally, we compute the reconstructed image, $\recIm$ as the first $\orgM$ rows and $\orgN$ columns of $\recPadIm = \DctMat{m}^{-1}\recF\DctMat{n}^{-\top} = \DctMat{m}^{\top}\recF\DctMat{n}$.

In the remainder of this section, we shall:
\begin{itemize}
    \item quantify how the success of \Alg\ref{alg:phaseRecovery} varies with the sampling frequency of the DCT,
    \item quantify how the distortion of the reconstructed image varies with the amount of pixels we solve the phase for (encoded by the parameter $\energyQuantile$ in \Sec\ref{sec:optAlg}),
    \item demonstrate the qualitative benefits of \Alg\ref{alg:greedyMwcs},
    \item demonstrate that the solve time for \Alg\ref{alg:phaseRecovery} grows linearly with respect to the number of pixels solved for given a sufficiently small $\energyQuantile$.
\end{itemize}

\begin{figure}[!t]
\centering
\includegraphics[width=0.47\textwidth]{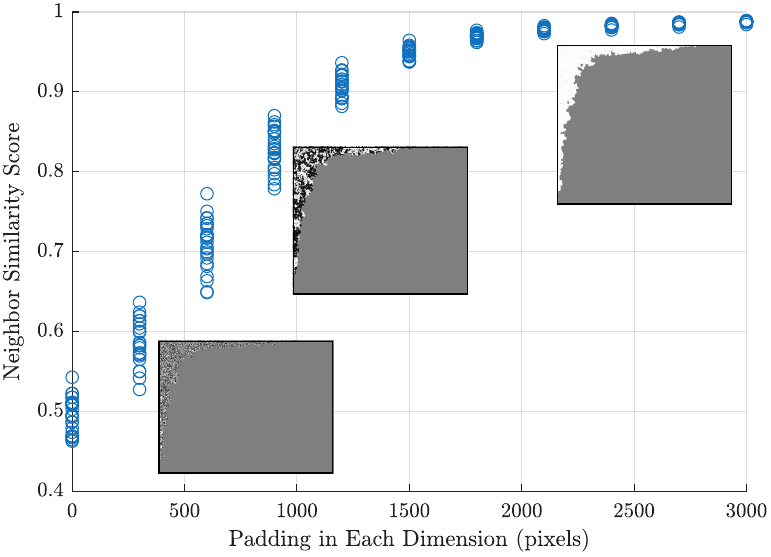}
\caption{Scatter plot of $\nss$ against $\padM=\padN$ for the phase retrieval of each image in the data set. A representative correctness matrix $\correctnessMat$ corresponding to one of the images in the data set is displayed immediately to the right of its padding value. In these images, white represents 1, gray represents 0 and black represents -1. As padding increases, $\nss$ increases and has less variance. The non-zero region of $\correctnessMat$ looks increasingly uniform as padding is added, indicating stronger performance.}
\label{fig:paddingVsSolveQuality}
\end{figure}

\subsection{Phase Retrieval Success vs Sampling Interval}\label{sec:padVsPRSuccess}
Given $\signSol$, the solution for the element-wise sign of $\disF$ given $\disG$, and $\disSign$, the true element-wise sign of $\disF$, let $\correctnessMat\in\{-1,0,1\}^{\imM\times\imN}$ be given by $\correctnessMat:=\signSol\odot\disSign$. That is, $\matElement{\correctnessMat}{\imIndI}{\imIndJ}$ is 1 if $\matElement{\signSol}{\imIndI}{\imIndJ}$ is correct, -1 if $\matElement{\signSol}{\imIndI}{\imIndJ}$ is incorrect and 0 if no solution was attempted at that element due to energy masking. Simply computing the ratio of elements of $\correctnessMat$ that are equal to 1 is not a good indicator for the success of the binary phase solution. For example, one could construct $\signSol$ by randomly choosing signs at each pixel. In expectation, the resulting $\correctnessMat$ would indicate a 1-to-1 ratio of correct to incorrect pixels, but success at any given pixel would be independent of success at neighboring pixels. As a comparison, if $\correctnessMat$ was such that $\matElement{\correctnessMat}{\imIndI}{\imIndJ} = 1$ if $\imIndI < \imM/2$ and $\matElement{\correctnessMat}{\imIndI}{\imIndJ} = -1$ otherwise, then $\correctnessMat$ would again indicate a 1-to-1 ratio of correct to incorrect pixels. However, in this later case, phase has been recovered almost perfectly, barring a sign change along a line. Loosely speaking, a large, continuous region of same-sign elements in $\correctnessMat$ indicates that the solver captured sign \textit{changes} consistent with those of $\disSign$. We adopt the following metric with this concept in mind.

Let $\neighborIndsSet_{\imInds}:= \{(\imIndI',\imIndJ')\in[\imM]\times [\imN]:\max(|\imIndI-\imIndI'|,|\imIndJ-\imIndJ'|) = 1\}$ denote the set of indices in the neighborhood of $(\imInds)$. Let $\validNeighs:=\{(\imInds)\in[\imM]\times [\imN]:\matElement{\correctnessMat}{\imIndI'}{\imIndJ'}\neq 0 \, \forall (\imIndI',\imIndJ')\in\neighborIndsSet_{\imInds}\}$ denote indices whose neighborhoods in the array $\correctnessMat$ have no zeros. We define the neighbor similarity score $\nss$ of the array $\correctnessMat$ as
\begin{equation}
    \nss := \frac{1}{|\validNeighs|}\sum_{(\imInds)\in\validNeighs}\frac{1}{|\neighborIndsSet_{\imInds}|}\sum_{(\imIndI',\imIndJ')\in\neighborIndsSet_{\imInds}}\equalIndicator(\matElement{\correctnessMat}{\imIndI}{\imIndJ},\matElement{\correctnessMat}{\imIndI'}{\imIndJ'}),
\end{equation}
where
\begin{equation}
    \equalIndicator(x,y) := \begin{cases}
        1, & x = y, \\
        0, & x\neq y.
    \end{cases}
\end{equation}
That is, the neighbor similarity score of $\correctnessMat$ computes the average over $(\imInds)\in\validNeighs$ of the ratio of elements of $\correctnessMat$ neighboring $\matElement{\correctnessMat}{\imIndI}{\imIndJ}$ that are equal to $\matElement{\correctnessMat}{\imIndI}{\imIndJ}$. Averaging over $\validNeighs$ ensures we ignore sections of $\correctnessMat$ that correspond to no attempt at computing phase. For low quality phase solutions, $\correctnessMat$ will appear to have random entries that have no correlation with their neighbors, resulting in $\nss$ near 0.5. As the quality of the phase retrieval solution improves, the value of an element of $\correctnessMat$ tends to match that of its neighbors. In this regime, $\nss$ approaches 1. 

\Fig\ref{fig:paddingVsSolveQuality} depicts the neighbor similarity score that results from running \Alg\ref{alg:phaseRecovery} on DCTs constructed from each image in the data set given some padding $\padM = \padN$ that was varied from 0 to 3000 in increments of 300. For this experiment, we set $\energyQuantile = 0.1$, $\bordWidth = 1$ and $\subblockM = \subblockN = 6$. As the padding increases and the cosine transforms are consequently upsampled, the neighbor similarity score increases and contains less variance, indicating better and more consistent performance across the data set. Upsampling the cosine transform enables the phase retrieval optimization to effectively detect discontinuities in the derivative of $\recF$, allowing the motivation presented in \Sec\ref{sec:hessNormMot} to hold.

\begin{figure*}[!t]
\centering
\includegraphics[width=0.9\textwidth,trim=0 0 0 4.5cm, clip]{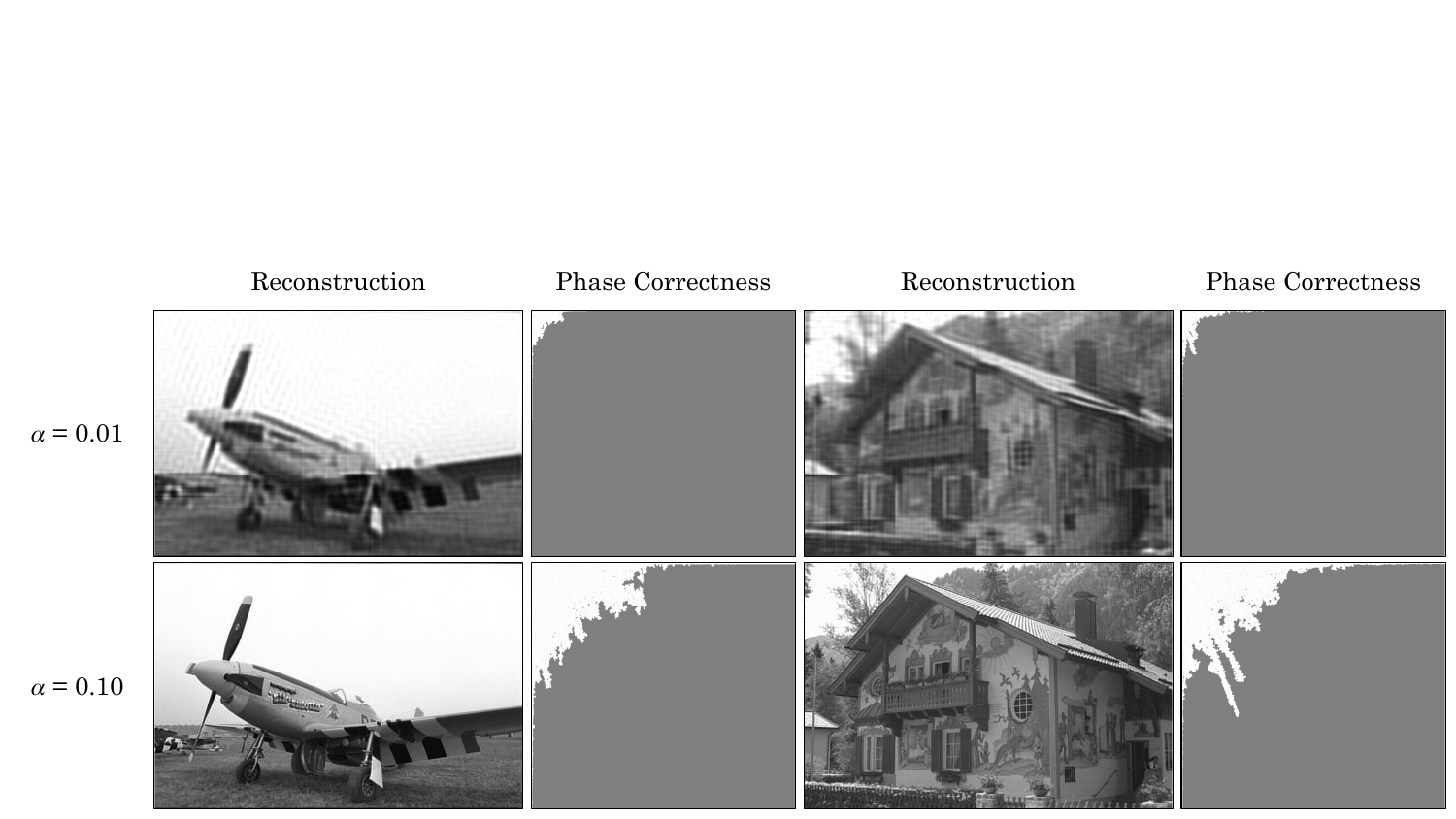}
\caption{Image reconstructions $\recIm$ for two images resulting from the experiment described in \Sec\ref{sec:brrVsDist} using $\energyQuantile = 0.01$ and $\energyQuantile = 0.1$. The ``phase correctness" displayed to the right of each reconstruction is the corresponding array $\correctnessMat$ as described in \Sec\ref{sec:padVsPRSuccess}. In the phase correctness images, white represents 1, gray represents 0 and black represents -1.}
\label{fig:brrVsQuality}
\end{figure*}

\begin{figure}[!t]
\centering
\includegraphics[width=0.47\textwidth]{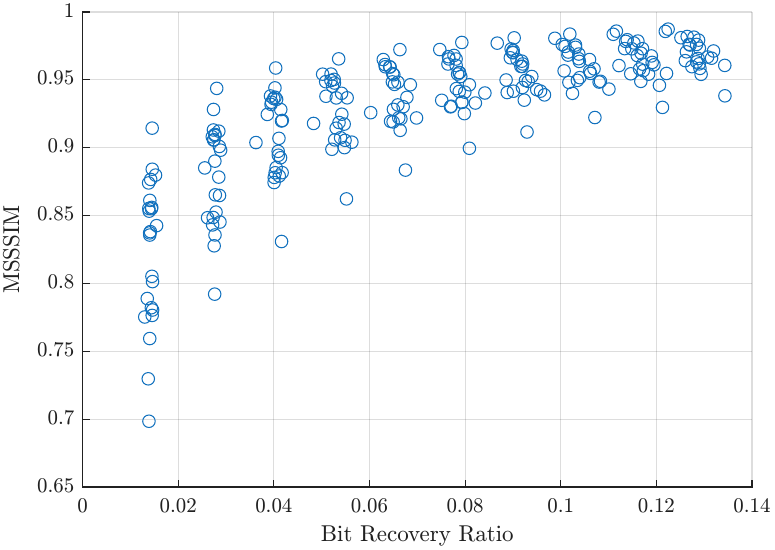}
\caption{Scatter plot of MSSSIM against bit recovery ratio. This plot illustrates how the distortion of a reconstructed image decreases as one solves the phase of more pixels. Reasonable reconstruction quality can be achieved by only solving the phase for a fraction of the total pixels. }
\label{fig:brrVsMsssim}
\end{figure}

\subsection{Pixels Solved vs Reconstruction Distortion} \label{sec:brrVsDist}
In order to quantify how the distortion between the original image, $\image$, and the reconstructed image, $\recIm$, varies as we vary the proportion of pixels we choose to solve phase over, we ran our experimental pipeline over the entire data set with the parameters $\padM = \padN = 3000$, $\bordWidth = 1$ and $\subblockM = \subblockN = 6$ while varying $\energyQuantile$ from 0.1 to 1.0 in increments of 0.1. We measured the distortion between $\image$ and $\recIm$ by computing their multiscale structural similarity (MSSSIM) index \cite{wang2003multiscale}. \Fig\ref{fig:brrVsMsssim} displays the results. The ``bit recovery ratio" indicates the number of pixels that phase was solved for divided by the number of elements ($\imM\imN$) of the cosine transform array. We used a large padding value, which ensured that phase was recovered accurately and consistently over the data set. The plot illustrates how the reconstruction quality improves as the proportion of pixels that we solve phase for increases. Image reconstructions achieved an MSSSIM of greater than 0.95 across almost the entire data set by setting $\alpha = 0.1$.

\Fig\ref{fig:brrVsQuality} depicts the qualitative results of \Alg\ref{alg:greedyMwcs} and \Alg\ref{alg:phaseRecovery}. As seen, low values of $\energyQuantile$ can result in blurring or ringing artifacts in the reconstruction of an image. These result from the loss of high frequency information in the phase solving process, as \Alg\ref{alg:greedyMwcs} masks these low-energy regions. The phase correctness plot for the image of the house highlights how \Alg\ref{alg:greedyMwcs} can highlight directional frequency bands that contain higher energy relative to the rest of the cosine transform.

\subsection{Solution Time}
\begin{figure}[!t]
\centering
\includegraphics[width=0.47\textwidth]{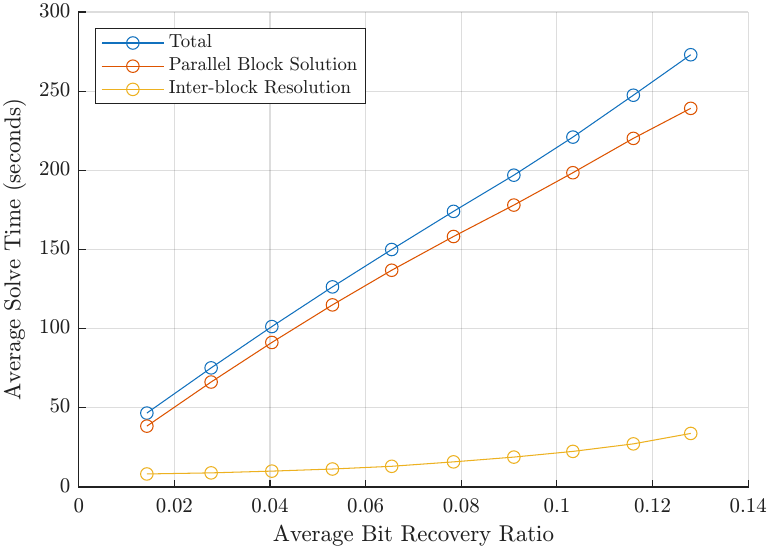}
\caption{Plot of phase solution time against the number of pixels solved. The parallel solution of individual blocks drives the total solution time in this regime and grows linearly with bit recovery ratio.}
\label{fig:brrVsTime}
\end{figure}

The experiment described in \Sec\ref{sec:brrVsDist} was run in the environment summarized in \Tab\ref{tab:hardware_software_specs} using parallel block processing with 12 workers. For each value of $\energyQuantile$, we measured the average across all images in the dataset of the amount of time it took to solve for the phase of all individual blocks in parallel (line \ref{alg:line:parPhaseSolve} in \Alg\ref{alg:phaseRecovery}) and the average amount of time it took to resolve inter-block sign discrepancies (line \ref{alg:line:blockResolve} in \Alg\ref{alg:phaseRecovery}). The results are displayed in \Fig\ref{fig:brrVsTime}. The amount of time it takes to solve all individual blocks in parallel grows linearly with the number of bits solved due to the fact that the size of individual blocks is constant. The amount of time it takes to resolve inter-block discrepancies is superlinear with respect to the number of bits solved due to the fact that the size of the feasible space in the inter-block resolution problem grows exponentially with the number of bits solved. However, over the range of $\energyQuantile$ that we tested, the inter-block resolution time was always significantly less than the individual block solution time, resulting in a total solve time that grows approximately linearly in this range.

\begin{table*}[!t]
\centering
\caption{Hardware and Software Environment}
\label{tab:hardware_software_specs}
\begin{tabular}{lll}
\toprule
\textbf{Category} & \textbf{Component} & \textbf{Specification} \\
\midrule
\textbf{Hardware} 
 & CPU & AMD Ryzen 9 7950X (16 Cores, 32 Threads, up to 5.7 GHz) \\
 & RAM & 128 GB DDR5 4800 MT/s \\
 & Storage & 2 TB NVMe PCIe 4.0 $\times 4$ SSD \\
\midrule
\textbf{Software} 
 & Operating System & Ubuntu 22.04.5 LTS (Linux 6.8 Kernel) \\
 & Programming Language & MATLAB R2022b \\
 & Optimization Solver & Gurobi 13.0.2  \\
\bottomrule
\end{tabular}
\end{table*}

\section{Conclusion}
In this work, we motivated and designed an algorithm that retrieves phase for the cosine transform of images by minimizing the curvature of the reconstructed transform. Unlike general phase retrieval algorithms, our algorithm leverages the energy compaction of cosine transforms and their restricted phase domain. We demonstrated that the algorithm can consistently reconstruct images to within an MSSSIM of 0.95.

While this algorithm functions well on computationally computed cosine transforms, it does not yet perform well on optically generated cosine transforms. Misalignment, ambient light, vibration and other factors present within a lab environment can introduce ``noise" into the cosine transform of an image, causing it to violate the motivation discussed in \Sec\ref{sec:hessNormMot}. In particular, the sharp curvature present along the zero level set of an image of an ideal cosine transform can become fairly smooth when non-ideal factors are introduced. Future work will explore how to make our algorithm robust to such issues.

\Alg\ref{alg:phaseRecovery} functions by decomposing a cosine transform into blocks, solving each block independently and then resolving sign discrepancies between each block. The subproblems are coupled together through the parameter $\bordWidth$ as well as the inter-block curvature. However, there is no guarantee that \Alg\ref{alg:phaseRecovery} converges to a global optimum of \eqref{eq:disHessMin}. Another direction for future work would involve employing an optimization technique that synergizes well with the divide-and-conquer nature of \Alg\ref{alg:phaseRecovery} while guaranteeing optimality, such as the alternating direction method of multipliers (ADMM). However, the iterative nature of ADMM may result in longer solve times.



{\appendices
\section{Proof of \Thm\ref{thm:uniqueness}}\label{appdx:uniquenessProof}
Let $\Fc := \cosTrans{f}$ and $\Gc := \cosTrans{g}$. Since $\f$ has compact support, $\Fc$ is an entire function in $\complex^2$ by the Paley-Wiener Theorem (see Theorem 5.4.10 in \cite{krantz2002primer}). Thus, its restriction to $\reals^2$ is real analytic. Under the same reasoning, $\Gc$ must also be real analytic. Manipulation of the condition $\Fc(\freqCoord)^2 = \Gc(\freqCoord)^2$ yields
\begin{equation}\label{eq:proofProd}
    (\Fc(\freqCoord) - \Gc(\freqCoord))(\Fc(\freqCoord) + \Gc(\freqCoord)) = 0
\end{equation}
for all $\freqCoord\in\reals^2$. Both of the factors in the equation above must be analytic since they are the sum of analytic functions. Suppose $\Fc(\freqCoord) - \Gc(\freqCoord)$ is nonzero at some $\freqCoord$. Then by the continuity of this term, it must be nonzero on a neighborhood of $\freqCoord$. In order for \eqref{eq:proofProd} to hold in that neighborhood, $\Fc(\freqCoord) + \Gc(\freqCoord)$ must be zero on the whole neighborhood. By the Identity Theorem for analytic functions, this would imply $\Fc(\freqCoord) + \Gc(\freqCoord)$ is identically zero on $\reals^2$. Swapping the two terms in the previous argument enables us to conclude that either $\Fc \equiv \Gc$  or $\Fc \equiv -\Gc$. Both $\f$ and $\g$ are assumed to be nonnegative, so $\Fc(\mathbf{0})$ and $\Gc(\mathbf{0})$ must both be positive. Thus, $\Fc \equiv \Gc$, and we have the desired result by the uniqueness of the cosine transform for even functions.

\section{Derivation of \eqref{eq:Trelation}}\label{sec:deriveB}
Suppose $\signBlkMat$ is an $\blockedM$ by $\blockedN$ array with columns $\signBlkVec_1$, $\signBlkVec_2$, ..., $\signBlkVec_\blockedN$. If one right-multiplies $\signBlkMat$ by a $\blockedN$ by $\subblockN$ matrix consisting of all zeros except for row $\imIndJ$ consisting of all ones, the resulting matrix consists of $\subblockN$ copies of $\signBlkVec_\imIndJ$ horizontally concatenated. Thus, to produce a matrix consisting of $\subblockN$ copies of $\signBlkVec_\imIndJ$ for all $\imIndJ\in[\blockedN]$ concatenated horizontally in order, one must right-multiply $\signBlkMat$ by $(\Id{\blockedN}\otimes\onesVec_{\subblockN})^\top$. Similarly, one can produce a matrix consisting of $\subblockM$ copies of each row of $\signBlkMat$ concatenated vertically in order by left-multiplying $\signBlkMat$ by $\Id{\blockedM}\otimes\onesVec_{\subblockM}$. Combining these two operations gives us
\begin{equation}
    \signBlkMat\otimes\onesVec_{\subblockM\times\subblockN} = (\Id{\blockedM}\otimes\onesVec_{\subblockM})\,\signBlkMat\,(\Id{\blockedN}\otimes\onesVec_{\subblockN})^\top.
\end{equation}
Applying the $\vectorize(\cdot)$ operation to both sides of this equation and utilizing \eqref{eq:vectorizeProduct} to simplify the right hand side yields the desired identity.

\section{Note on \Assump\ref{assump:noDegenCurve}}\label{appdx:nonDegeneracy}
A simple counterexample to \Assump\ref{assump:noDegenCurve} may be constructed. One need simply to convolve two images together and the cosine transform of the result will have degenerate continuous curves on which the transform and its gradient are simultaneously zero. We shall demonstrate, however, that such counterexamples are topologically rare. 

Let $\imageSet$ denote the set of even images as defined in \Sec\ref{sec:cosTransProps}. Given an $\f\in\imageSet$, let $\Fc:=\cosTrans{f}$. Let $\degenLocus{\f}$ denote the set of $\freqCoord\in\reals^2$ such that $\Fc(\freqCoord) = 0$ and $\nabla\Fc(\freqCoord) = \mathbf{0}$. Finally, given a compact domain $\contHessMinDom\subset\posReals^2$, denote by $\degenImageSet$ the set of $\f\in\imageSet$ such that $\degenLocus{\f}\cap\contHessMinDom \neq \emptyset$.

\begin{theorem}
    $\degenImageSet$ is nowhere dense in $\imageSet$. Furthermore, $\imageSet \setminus \degenImageSet$ is dense in $\imageSet$.
\end{theorem}
\begin{proof}
    We shall use $\Vert \cdot\Vert$ to denote the Euclidean norm of a vector and $\Vert\cdot\Vert_1$ to denote the $L^1$ norm of a function, the latter of which we use to induce a topology on $\imageSet$. Given an arbitrary $\f\in\imageSet$ and $\Fc:=\cosTrans{f}$, we construct an image perturbation $\pc\in\imageSet$ parameterized by $\pertParams\in\posReals^3$. Let $\coord_2 := (\shift, 0)^\top$ and $\coord_3 := (0, \shift)^\top$, where $a\in\posReals$. Then define
    \begin{align}
        \pertComp_1(\coord) &:= \delta(\coord), \\
        \pertComp_2(\coord) &:= \nicefrac{1}{2}(\pertComp_1(\coord - \coord_2) + \pertComp_1(\coord + \coord_2)), \\
        \pertComp_3(\coord) &:= \nicefrac{1}{2}(\pertComp_1(\coord - \coord_3) + \pertComp_1(\coord + \coord_3)),
    \end{align}
    and set $\pc := \sum_{k=1}^3 \pertParam_k \pertComp_k$,
    the cosine transform of which is denoted $\Pc$. The use of a Dirac delta in the construction of $\pc$ means it is not strictly an element of $\imageSet$, but one can construct a similar proof by substituting the Dirac delta with the function 
    \begin{equation}
    \pertComp_1(\coord) := \begin{cases}
        \radius^{-2}, & \Vert \coord \Vert < \radius, \\
        0, & \Vert \coord \Vert \geq \radius,
        \end{cases}
    \end{equation}
    and selecting $\radius\in\posReals$ sufficiently small. The resulting perturbed image is in $\imageSet$. We shall continue with our Dirac delta formulation for ease of exposition.

    Let $\jet:\contHessMinDom\times\posReals^3\mapsto\reals^3$ denote the map from frequency coordinates $\freqCoord$ and parameter coordinates $\pertParams$ to the 1-jet of the perturbed image's cosine transform:
    \begin{equation}
        \jet(\freqCoord;\pertParams) := \begin{pmatrix}
            \Fc(\freqCoord) + \Pc(\freqCoord) \\
            \nabla\Fc(\freqCoord) + \nabla\Pc(\freqCoord)
        \end{pmatrix}.
    \end{equation}
    A point $\freqCoord$ is an element of $\degenLocus{\f+\pc}$ exactly when $\jet(\freqCoord;\pertParams) = \mathbf{0}$. Computing the gradient of $\jet$ with respect to $\pertParams$ yields
    \begin{equation}
        \nabla_\pertParams \jet(\freqCoord) = 
        \begin{pmatrix}
            1 & \cos(2\pi\shift\omega_1) & \cos(2\pi\shift\omega_2) \\
            0 & -2\pi\shift\sin(2\pi\shift\omega_1) & 0 \\
            0 & 0 & -2\pi\shift\sin(2\pi\shift\omega_2)
        \end{pmatrix},
    \end{equation}
    which is nonsingular for all $\freqCoord\in\contHessMinDom$ given a sufficiently small $\shift$. Thus, $\nabla_{\freqCoord,\pertParams} \jet (\freqCoord;\pertParams)$ has full row rank and $\jet$ is consequently a submersion. Let $\jetc:\contHessMinDom\mapsto\reals^3$ denote the restriction of $\jet (\freqCoord;\pertParams)$ to a particular $\pertParams$. By the Parametric Transversality Theorem (see Theorem 6.35 in \cite{lee2013introduction}) $\jetc$ is transverse to the origin for almost all $\pertParams$. Thus, we can choose $\pertParams$ such that the perturbed image $\f+\pc$ is arbitrarily close to the original image $\f$ in the $L^1$ norm and $\jetc$ is transverse to the origin in $\reals^3$. 
    
    We have shown that the set of images with a 1-jet transverse to the origin in $\reals^3$ is dense in $\imageSet$. However, the codimension of the origin in $\reals^3$ is 3, which is larger than the dimension of $\contHessMinDom$, which is 2. Transversality of $\jetc$ to the origin dictates that this is only possible if $\jetc(\contHessMinDom)$ fails to intersect the origin (see Theorem 6.30 of \cite{lee2013introduction}). Thus, $\imageSet \setminus \degenImageSet$ is dense in $\imageSet$.

    In order to show that $\degenImageSet$ is nowhere dense in $\imageSet$, we only have left to show that $\imageSet \setminus \degenImageSet$ is open, as the complement of a dense, open set is nowhere dense. Suppose $\f\in\imageSet \setminus \degenImageSet$. Then $\Fc$ and $\nabla\Fc$ are continuous (we demonstrated in \Appdx\ref{appdx:uniquenessProof} that $\Fc$ was analytic). Since $\degenLocus{\f}\cap\contHessMinDom = \emptyset$ and $\contHessMinDom$ is compact by assumption, we must have that $\Vert (\Fc(\freqCoord), \nabla\Fc(\freqCoord)^\top)^\top\Vert$ attains a minimum value $\varepsilon>0$ over $\contHessMinDom$. It is straightforward to show 
    \begin{equation}
        \left\Vert \begin{pmatrix}
            \Fc(\freqCoord) \\
            \nabla\Fc(\freqCoord)
        \end{pmatrix}\right\Vert \leq \sqrt{1+C}\Vert \f \Vert_1,
    \end{equation}
    where $C>0$ depends on the radius of the support of functions in $\imageSet$. Now, consider the open ball of functions $g\in\imageSet$ such that $\Vert f -g\Vert_1 <  \varepsilon/(2\sqrt{1+C})$. Then, by the triangle inequality, we have that for all $\freqCoord\in\contHessMinDom$,
    \begin{align}
        \left\Vert \begin{pmatrix}
            \Gc(\freqCoord) \\
            \nabla\Gc(\freqCoord)
        \end{pmatrix}\right\Vert &\geq \left\Vert \begin{pmatrix}
            \Fc(\freqCoord) \\
            \nabla\Fc(\freqCoord)
        \end{pmatrix}\right\Vert - \left\Vert \begin{pmatrix}
            \Fc(\freqCoord) - \Gc(\freqCoord)\\
            \nabla\Fc(\freqCoord) - \nabla\Gc(\freqCoord)
        \end{pmatrix}\right\Vert \\
        &\geq \varepsilon - \sqrt{1+C}\Vert f-g\Vert_1 \geq \varepsilon/2.
    \end{align}
    That is, for all $\freqCoord\in\contHessMinDom$, $\freqCoord\notin\degenLocus{g}$. Thus, all such $g$ are elements of $\imageSet\setminus\degenImageSet$ and we are done.

\end{proof}

}

\bibliographystyle{IEEEtranN}
\bibliography{references}

\newpage

\begin{IEEEbiography}[{\includegraphics[width=1in,height=1.25in,clip,keepaspectratio]{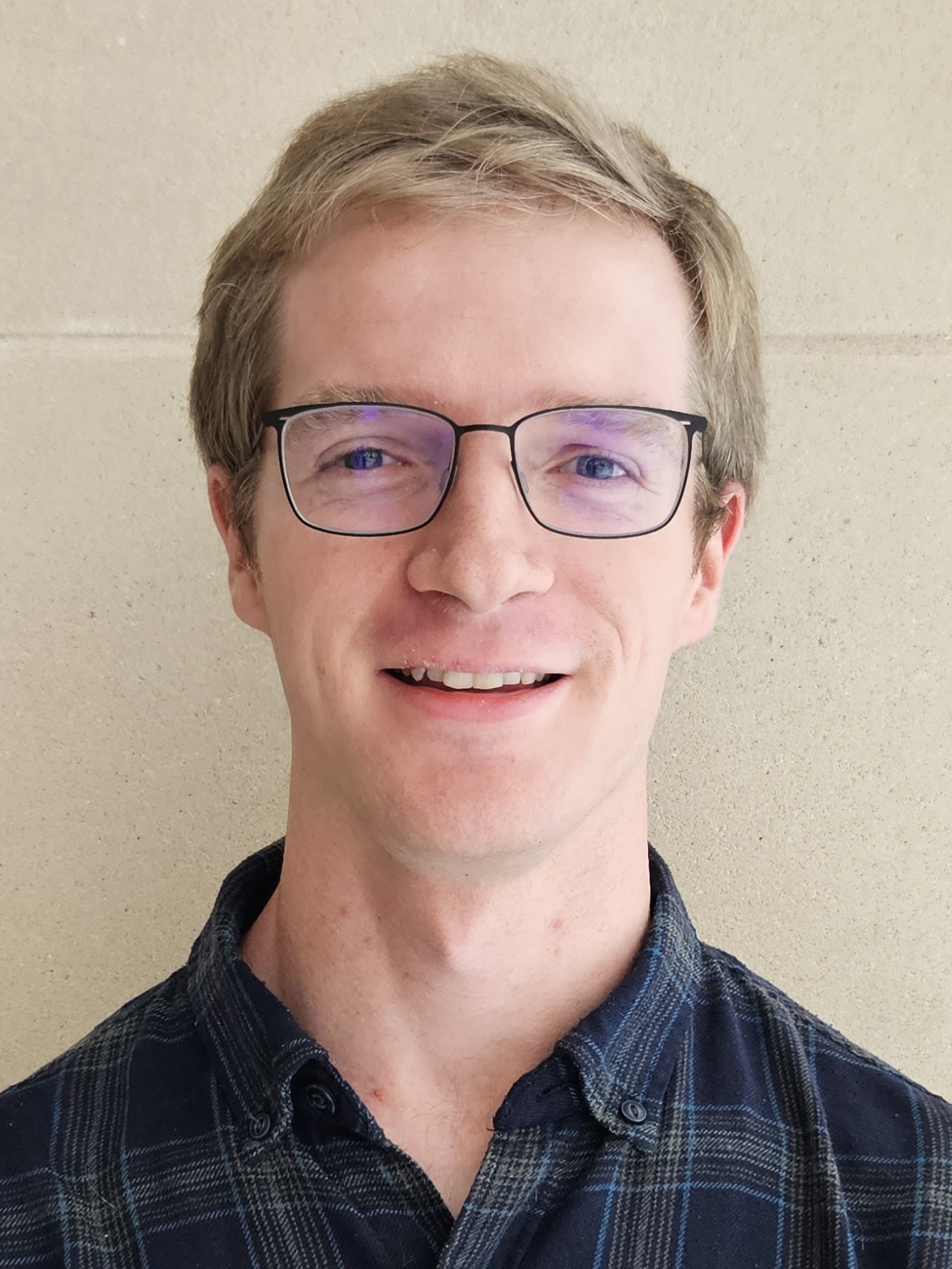}}]{Ronald Ogden} (Student Member, IEEE) received the B.S. degree in Aerospace Engineering and Mathematics from the Massachusetts Institute of Technology in 2018. He was a flight test engineer at Wisk Aero, a company developing autonomous aircraft for urban air mobility. Now, he is a Ph.D. candidate in the Department of Aerospace Engineering and Engineering Mechanics at the University of Texas at Austin. He currently researches how to leverage optics for real-time vision compression in robotics systems.
\end{IEEEbiography}

\begin{IEEEbiography}[{\includegraphics[width=1in,height=1.25in,clip,keepaspectratio]{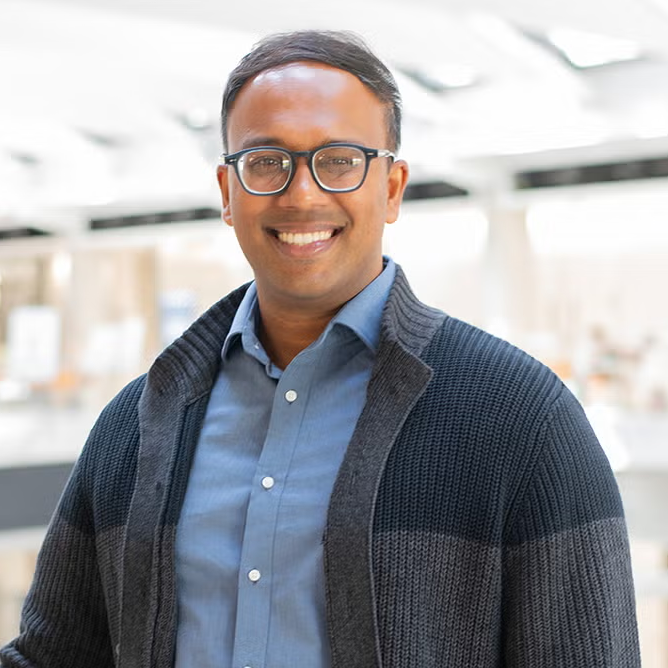}}]{Shwetadwip Chowdhury} (Member, IEEE) received the B.S. and Ph.D. degrees both in biomedical engineering from Duke University. Afterwards, he was a Postdoctoral Associate in the Department of Electrical Engineering and Computer Science at the University of California, Berkeley. He is now an Assistant Professor in the Department of Electrical and Computer Engineering at the University of Texas at Austin. Chowdhury is the recipient of an NSF
CAREER Award.
\end{IEEEbiography}

\begin{IEEEbiography}[{\includegraphics[width=1in,height=1.25in,clip,keepaspectratio]{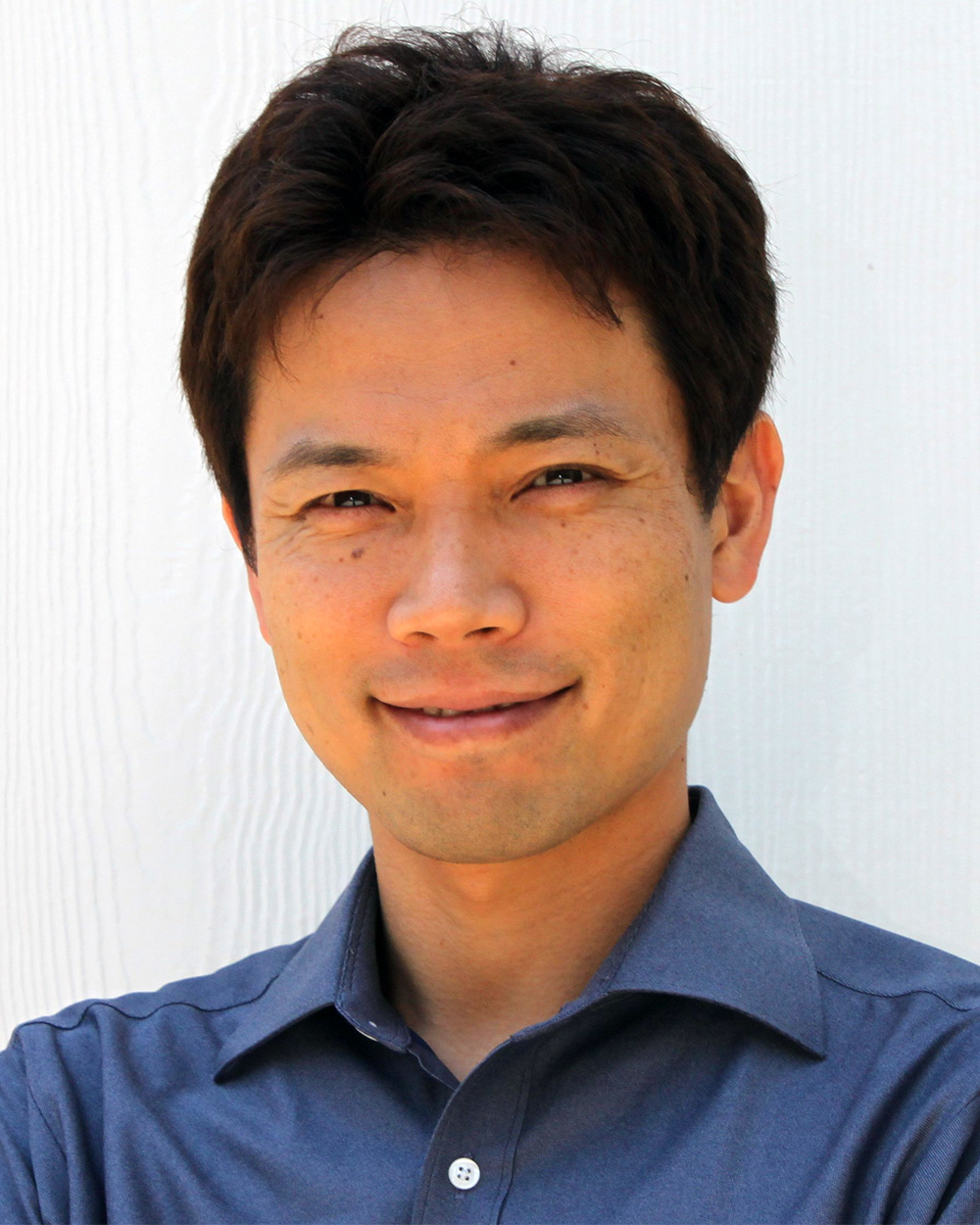}}]{Takashi Tanaka} (Senior Member, IEEE)
received the B.S. degree from the
University of Tokyo, Tokyo, Japan, in 2006, and the
M.S. and Ph.D. degrees in aerospace engineering
(automatic control) from the University of Illinois
at Urbana Champaign, Champaign, IL, USA, in 2009 and 2012, respectively. He was a Postdoctoral
Associate with the Laboratory for Information and
Decision Systems at the Massachusetts Institute of
Technology, Cambridge, MA, USA, from 2012 to
2015, and a postdoctoral researcher at KTH Royal
Institute of Technology, Stockholm, Sweden, from
2015 to 2017. He was an Assistant Professor in the Department of Aerospace
Engineering and Engineering Mechanics at the University of Texas at Austin
between 2017 and 2024, and is an Associate Professor at the School of Aeronautics and Astronautics and the Elmore Family School of Electrical and Computer Engineering at Purdue University since 2025. 
He was the recipient of the DARPA Young Faculty Award, the AFOSR Young Investigator Program Award, and the NSF Career Award.
\end{IEEEbiography}

\begin{IEEEbiography}[{\includegraphics[width=1in,height=1.25in,clip,keepaspectratio]{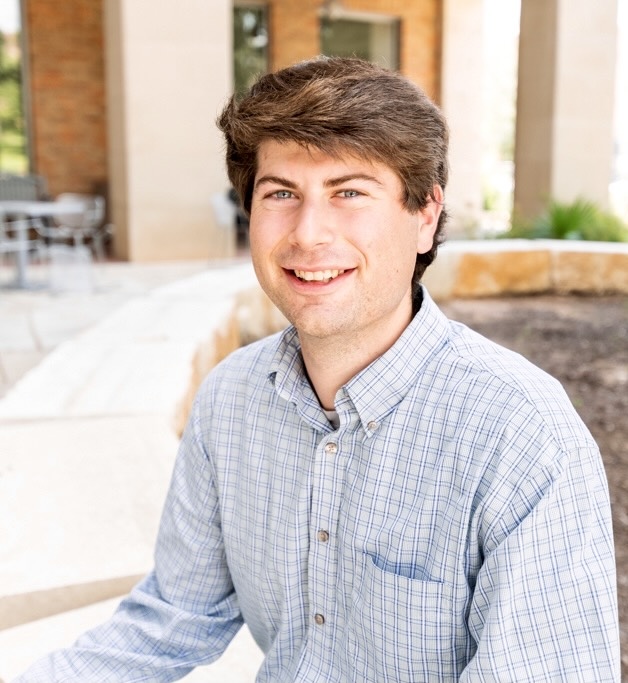}}]{David Fridovich-Keil} (Senior Member, IEEE) received the B.S.E. degree in electrical engineering from Princeton University, and the Ph.D.
degree from the University of California, Berkeley. He is an Assistant Professor in the Department of Aerospace Engineering and Engineering Mechanics at the University of Texas at Austin. Fridovich-Keil is the recipient of an
NSF Graduate Research Fellowship and an NSF
CAREER Award.
\end{IEEEbiography}

\vfill

\end{document}